\documentclass[aps,prd,10pt,twocolumn,superscriptaddress,showpacs]{revtex4-1}
\pdfoutput=1
\usepackage{hyperref}
\usepackage{graphicx}
\usepackage{enumerate}
\usepackage{amsfonts,amsmath,amssymb,bm,bbm}
\usepackage{color}
\usepackage{epsfig, subfigure}
\usepackage{setspace}
\usepackage{booktabs, tabularx} 

\hypersetup{
    pdfnewwindow=true,      % links in new window
    colorlinks=true,       % false: boxed links; true: colored links
    linkcolor=blue,          % color of internal links
    citecolor=blue,        % color of links to bibliography
    filecolor=blue,      % color of file links
    urlcolor=blue        % color of external links
}

\newcommand\blue{\color{blue}}

\newcommand{\be}{\begin{equation}}  
\newcommand{\ee}{\end{equation}}
\newcommand{\bea}{\begin{eqnarray}}  
\newcommand{\eea}{\end{eqnarray}}

\newcommand{\sigv}{{\langle\sigma v\rangle}}

\begin{document}

\title{\mbox{\hspace{-0.25cm}Galactic Center Radio Constraints on Gamma-Ray Lines from Dark Matter Annihilation}}

\author{Ranjan Laha}
\email{laha.1@osu.edu}
\affiliation{{\mbox{Center for Cosmology and AstroParticle Physics, Ohio State University, Columbus, OH 43210, USA.}}}
\affiliation{{Department of Physics, Ohio State University, Columbus, OH 43210, USA.}}

\author{Kenny Chun Yu Ng}
\email{ng.199@osu.edu}
\affiliation{{\mbox{Center for Cosmology and AstroParticle Physics, Ohio State University, Columbus, OH 43210, USA.}}}
\affiliation{{Department of Physics, Ohio State University, Columbus, OH 43210, USA.}}

\author{Basudeb Dasgupta}
\email{dasgupta.10@osu.edu}
\affiliation{{\mbox{Center for Cosmology and AstroParticle Physics, Ohio State University, Columbus, OH 43210, USA.}}}

\author{Shunsaku Horiuchi}
\email{horiuchi@mps.ohio-state.edu}
\affiliation{{\mbox{Center for Cosmology and AstroParticle Physics, Ohio State University, Columbus, OH 43210, USA.}}}

\date{\today}

%%%%%%%%%%%%%%%%%%%%%%%%%%%%%%%%%%%%%%%%%%
\begin{abstract}
Recent evidence for one or more gamma-ray lines at $\sim130$\,GeV in the Fermi-LAT data from the Galactic Center has been interpreted as a hint for dark matter annihilation to $Z\gamma$ or $H\gamma$ with an annihilation cross section of $\langle \sigma v\rangle$ $\sim10^{-27}{\rm cm^3\,s^{-1}}$. We test this hypothesis by comparing synchrotron fluxes due to the electrons and positrons from decay of the $Z$ or the $H$ bosons {\it only} against radio data from the same region in the Galactic Center. We find that the radio data from single-dish telescopes marginally constrain this interpretation of the claimed gamma lines for a contracted NFW profile. Already-operational radio telescopes, such as LWA, VLA-Low and LOFAR, and future radio telescopes like SKA, are sensitive to annihilation cross sections of the order of $10^{-28}{\rm cm^3\,s^{-1}}$, and can confirm or rule out this scenario very soon. We discuss the dependencies on the dark matter profile, magnetic fields, and background radiation density profiles, and show that the constraints are relatively robust for any reasonable assumptions. Independent of the above said recent developments, we emphasize that our radio constraints apply to all models where dark matter annihilates to $Z\gamma$ or $H\gamma$.

\end{abstract}
%%%%%%%%%%%%%%%%%%%%%%%%%%%%%%%%%%%%%%%%%%
%Dark Matter 

%\preprint{}

\maketitle

%%%%%%%%%%%%%%%%%%%%%%%%%%%%%%%%%%%%%%%%%%
\section{Introduction}
\label{sec:introduction}
%%%%%%%%%%%%%%%%%%%%%%%%%%%%%%%%%%%%%%%%%%
The particle identity of dark matter (DM) is one of the outstanding puzzles in contemporary physics. In order to fully understand the particle properties of dark matter, a number of complementary approaches to dark matter searches have been adopted. Indirect detection of dark matter is a promising technique, in which the products of dark matter annihilation are searched for, and gives us information about the DM abundance and annihilation rate at various astrophysical sites~\cite{Jungman:1995df, Bertone:2004pz, Bergstrom:2012np,Feng:2010gw,Gunn:1978gr,Zeldovich:1980st}.

Gamma-ray lines from DM self annihilation are believed to be a smoking-gun signature, and have been investigated in considerable detail~\cite{Bergstrom:1988fp, Flores:1989ru, Bergstrom:1997fh, Bern:1997ng, Ullio:1997ke, Bertone:2009cb}. Despite the relative freedom in DM model-building, if DM self-annihilation is to two-body Standard Model final states, then gamma-ray line(s) can be produced only via the following three channels{\blue :} (i) $\chi\chi \rightarrow \gamma \gamma$, (ii) $\chi\chi \rightarrow Z \gamma$, and (iii) $\chi\chi \rightarrow H \gamma$, where $\chi$ denotes the DM, and $Z$ and $H$ denote the $Z$ and Higgs boson, respectively. We take the mass of the Higgs boson to be 125\,GeV~\cite{:2012gu, :2012gk}, and allow a heavy DM to annihilate to it.

Recently, evidence for a gamma-ray line from the Galactic Center (GC) has been uncovered in the Fermi-LAT data at $\sim130$ GeV~\cite{Weniger:2012tx, Bringmann:2012vr} and this has given rise to renewed interest in considering the line signal in more detail~\cite{Tempel:2012ey,Boyarsky:2012ca, Rajaraman:2012db, GeringerSameth:2012sr, Su:2012ft, Buchmuller:2012rc, Cohen:2012me, Cholis:2012fb, Su:2012zg, Bergstrom:2012vd, Hektor:2012kc,Yang:2012ha,Huang:2012yf,Feng:2012gs,Li:2012qg,Whiteson:2012hr}. 

This statistically significant signal has been tentatively interpreted as arising from DM annihilation. Generally speaking, the signal requires a DM self annihilation cross section of $\sigv\sim10^{-27}\,\rm{cm^3\,s^{-1}}$ and the Galactic DM halo described by a standard NFW, Einasto, or a contracted NFW profile. Subsequently, a variety of particle physics models have been proposed to explain the signal~\cite{Ibarra:2012dw, Dudas:2012pb, Cline:2012nw, Choi:2012ap, Kyae:2012vi, Lee:2012bq, Acharya:2012dz, Buckley:2012ws, Chu:2012qy, Das:2012ys, Kang:2012bq, Weiner:2012cb, Heo:2012dk,Tulin:2012uq,Li:2012jf,Park:2012xq,Frandsen:2012db,Oda:2012fy,Cline:2012bz,Bai2012,Shakya:2012fj,Bergstrom:2012bd,Fan:2012gr}. It is also found that the line is off-center from the GC by approximately $1.5^\circ$~\cite{Su:2012ft, Yang:2012ha}, which requires the center of the DM halo to be displaced from the baryonic center. This degree of displacement appears reasonable as shown by recent numerical simulations~\cite{Kuhlen:2012qw}.

On the other hand, there are arguments against the DM origin of the gamma-ray line. There are hints that the line is also present in the photons collected from the cascades in the Earth's atmosphere, which is a ``pure background" region~\cite{Su:2012ft}, although this has been claimed to be due to statistical fluctuations~\cite{Finkbeiner:2012ez,Hektor:2012ev}. There have been claims of the presence of gamma-ray lines at the same energy, spatially correlated with some Fermi-LAT unassociated sources~\cite{Su:2012zg}. However, there are also counterclaims that most of these unassociated sources are consistent with being standard astrophysical objects such as active galactic nuclei or statistical fluctuations~\cite{Hooper:2012qc,Mirabal:2012za,Hektor:2012jc}. Furthermore, it remains possible that the GC line signal is also of an astrophysical origin~\cite{Boyarsky:2012ca,Profumo:2012tr, Aharonian:2012cs}. 

The Fermi-LAT collaboration, in their search for $\gamma$-ray lines in the 2 year data set~\cite{Ackermann:2012qk} did not find a signal as the analysis employed a  different search strategy, an older data set and background rejection software, and a larger search region, making it difficult to compare directly with the above claims. However in their most recent search for gamma-ray lines with the 4 year data~\cite{Albert:Fermi2012}, the Fermi-LAT collaboration has acknowledged the presence of a feature at the GC at 135 GeV (this shift in the energy is due to recalibration but we will assume that the line is at 130 GeV throughout this work). The collaboration also finds a feature at the Earth limb at the same energy~\cite{Charles:Fermi2012}. The collaboration states that it does not have a consistent interpretation of the Galactic Center feature and that it needs more data to resolve the issue~\cite{Bloom:Fermi2012}.
Given the arguments in favor of and against the DM origin of this signal, this remains a topic of active research.

If the DM annihilates to two-body Standard Model final states, as in (i)-(iii), then we can predict some particle physics model-independent consequences. For a dominant annihilation (i), i.e., to two photons, there are no further interactions of the photons at an appreciable level, with all higher-level amplitudes suppressed by at least $\alpha~\approx~1/137$. However, if the annihilation proceeds as~in~(ii) or (iii), i.e., to a photon and a heavy Standard Model boson, the heavy boson decays to other charged particles which can have observable consequences.

The decay of the $Z$ and the $H$ boson produces electrons, protons, neutrons, neutrinos, their antiparticles, as well as photons as final states. The almost featureless spectra of these secondary particles poses considerable difficulty in their search above the astrophysical backgrounds. Searches for antimatter benefit from lower cosmic ray backgrounds, therefore one can search for antiprotons and positrons from the $Z$ and the $H$ boson. A search for antiprotons from these decays constrains several particle physics models which can give rise to a gamma-ray line~\cite{Buchmuller:2012rc}, whereas the preexisting unaccounted excess in positrons~\cite{Adriani:2008zr, FermiLAT:2011ab} makes a positron search ambiguous. Neutrinos could, in principle, be used to distinguish between all three final states, but achieved or projected sensitivities in the range \mbox{$\sigv\sim(10^{-22}\,{\rm cm^3\,s^{-1}} - 10^{-23}\,{\rm cm^3\,s^{-1}}$)~\cite{Abbasi:2011eq,Beacom:2006tt,Yuksel:2007ac,Dasgupta:2012bd}} will not be able to probe the claimed signal. Secondary photons that are produced in the decay of the $Z$ or the $H$ boson, or in other DM annihilation channels, also constrain these scenarios~\mbox{\cite{Buchmuller:2012rc, Cohen:2012me, Cholis:2012fb}}, and there are ongoing efforts to confirm this 130\,GeV line with future detectors~\cite{Bergstrom:2012vd,Li:2012qg}.

In this paper, we ask the questions -- If the 130\,GeV signal is indeed from DM annihilation to $Z \gamma$ or $H \gamma$, what other consequences are guaranteed? Can we use these consequences to test this signal? Synchrotron radiation from products of DM annihilation has been argued to provide strong constraints for many DM annihilation channels and scenarios~\cite{Aloisio:2004hy, Borriello:2008gy, Bertone:2008xr, Ishiwata:2008qy, Fornengo:2011iq, Crocker:2010gy, Gondolo:2000pn, Bergstrom:2008ag, Hooper:2007kb, Regis:2008ij, Bertone:2001jv, Hooper:2010im, Hooper:2012jc, Fornengo:2011cn, Zhang:2009pr, Zhang:2008rs,Mambrini:2012ue,Fornengo:2011xk}. Thus, following these promising leads, we explore our question by calculating the synchrotron radiation in the GC due to the electrons and positrons from $Z$ or $H$ decays {\it only}, and comparing it to existing data from radio telescopes.

We first take a very conservative approach, where we compare the DM-induced synchrotron fluxes to the total measured radio flux at 330\,MHz in a relatively large region around the GC, and determine that DM annihilation cross sections to these channels cannot be more than $\sigv\sim10^{-25}{\rm cm^3\,s^{-1}}$. However, this approach is overly conservative, as the synchrotron fluxes in the GC are modeled accurately with known astrophysics. We argue that the flux due to dark matter must not exceed the uncertainties on the modeled radio fluxes, which provides us with a constraint $\sigv \sim10^{-26}{\rm cm^3\,s^{-1}}$. Constraints obtained by comparing fluxes predicted in smaller regions of interest and upper limits at 408\,MHz imply $\sigv\sim10^{-27}{\rm cm^3\,s^{-1}}$, and are already in mild tension with the 130\,GeV line. We forecast that the sensitivity can be improved to $\sim10^{-28}{\rm cm^3\,s^{-1}}$ with a few hours of observation of the GC at 80 MHz with LWA, the Long Wavelength Array, and at 200\,MHz with LOFAR, the LOw-Frequency ARray for radio astronomy, allowing us to constrain interpretations of the 130\,GeV line signal in the Fermi-LAT data in terms of DM annihilation to $Z \gamma$ or $H \gamma$. Although this is the main motivation for our present work, the radio constraints we derive are valid regardless of whether this claimed 130\,GeV line signal survives further scrutiny or not. These constraints will continue to apply to any future interpretations of gamma-ray lines at the GC in terms of DM annihilation.

Note that, these sensitivities readily probe the cross section that explains the tentative 130\,GeV line signal. More generally, we expect these sensitivities to be able to probe many of the models, not necessarily supersymmetric, that could explain this signal. We also emphasize that since we are looking for the synchrotron radiation from the electrons and positrons produced in the decays of the $Z$ or the $H$ boson {\it only}, our constraints are independent of the underlying DM particle physics model. In these two ways, our work is complementary to Ref.\,\cite{Cohen:2012me}. The results here are of course affected by astrophysical uncertainties, e.g., dark matter density profile, magnetic fields, interstellar radiation energy density, and proton density in the Galaxy, and by taking a range of different values for them we try to understand their impact.
 
The rest of the paper is organized as follows. In Sec.\,\ref{sec:DataTheory} we discuss the radio data that we use for obtaining our constraints, and the theoretical framework for calculating the flux densities from synchrotron radiation by DM annihilation products. In Sec.\,\ref{sec:Inputs} we furnish and justify the astrophysical inputs, i.e., DM density, magnetic fields, and radiation density in the Galaxy, that we use for our calculations. In Sec.\,\ref{sec:results} we show the predicted flux densities for benchmark DM annihilation cross sections, and provide constraints on the DM annihilation cross section as a function of DM mass for the channels \mbox{$\chi\chi\rightarrow Z\gamma$} and $\chi\chi\rightarrow H\gamma$, and conclude in Sec.\,\ref{sec:outlook}.

\section{Experimental Data and Theoretical Framework}
\label{sec:DataTheory}

\subsection{Radio data and regions of interest}

We use radio data at two frequencies, 330\,MHz and 408\,MHz,  to obtain the limits on DM self annihilation cross section from near the GC. We use the projected radio sensitivity of the LWA telescope at 80\,MHz~\cite{Kassim} to predict the future sensitivity on DM self annihilation cross section to gamma-ray lines that can be probed by radio data. We calculate the synchrotron flux density at a region offset from the GC in the 200\,MHz band, which is an operating frequency band for the LOFAR telescope.

We consider both the LWA and LOFAR telescopes for two reasons. First, the geographical location of LOFAR is not ideal to observe the GC, but it can measure the radio flux away from the GC to derive useful constraints on dark matter properties. Second, LWA is in a much better location to study the GC, but the GC can be opaque at the frequencies the LWA will operate in. The redeeming factor is that the absorptive nature of the GC at these frequencies is patchy and there exists regions that are transparent~\cite{2003ANS...324Q..65K,1985ApJ...299L..13L,1986Natur.322..522K,2003ANS...324...17B}. This argues for region optimization depending on the observed patchiness and sizes.

\subsubsection{ROI-2$^\circ$: Region of interest 2$^\circ$ around GC}
The radio measurements in the 330\,MHz band by the Green Bank telescope are available in a 6$^\circ$ $\times$ 2$^\circ$ region around the GC~\cite{2010Natur.463...65C}, and provides us our first region of interest (ROI-2$^\circ$). We approximate this region to be a circle with a radius~0.034 radians ($\approx$ 2$^\circ$), for simplicity. Thus, we approximately match the area of the region of observation in Ref.\,\cite{2010Natur.463...65C}, but the shape is different. We assume that this difference will not change our results significantly. In Ref.\,\cite{2010Natur.463...65C}, the authors present an astrophysical model to explain the data, so we use the uncertainty in their measurement at 330\,MHz, i.e., $0.05\times18000$\,Jy = 900\,Jy, to obtain our limits on the self annihilation cross section as a function of the DM mass. The authors in Ref.\cite{2010Natur.463...65C} also use radio data at higher frequencies to construct a GC model. Comparing our calculated synchrotron flux density with the errors in their measurement at every measured frequency we find that the most constraining limit on DM self annihilation cross section comes from the lowest frequency band (330\,MHz) and hence we only use the uncertainty in the measurement at 330\,MHz to constrain DM properties.

We will also use this ROI to obtain our projected sensitivity on DM particle properties using the future measurement around the GC by the LWA telescope.

\subsubsection{ROI-4$''$: Region of interest 4$''$ around GC}
The radio measurement in the 408\,MHz band by the Jodrell Bank telescope ~\cite{1976MNRAS.177..319D} in a region 4$''$ around the GC provides us with our second region of interest \mbox{(ROI-4$''$)}. At this frequency, the region of interest is a circle of angular radius 4$''$ and the upper limit on the radio flux is 50\,mJy. This region is significantly smaller than ROI-2$^\circ$, and as we will show, is affected differently to our input parameters. Thus it provides a complementary site to testing DM properties.

\subsubsection{ROI-away: Region of interest away from the GC}

We also calculate the synchrotron flux within an angular cone of radius 1$^\circ$, at angles 1.5$^\circ$ and 10$^\circ$ away from the GC.
We calculate the synchrotron flux at regions away from the GC at 80\,MHz, which is an operating frequency band for the LWA telescope, and at 200\,MHz, which is an operating frequency band for the LOFAR telescope. We calculate how the synchrotron flux varies with mass of the DM, for a given $\langle \sigma v \rangle$.
%, at an angle 1.5$^\circ$ and 10$^\circ$ away from the GC and within an angular cone of radius 1$^\circ$. We will also show our calculated synchrotron flux, for a given DM mass and $\langle \sigma v \rangle$, against angle away from the GC in a circular region of radius 1$^\circ$. 
The advantage of measuring the synchrotron flux away from the GC is that the synchrotron flux depends less on the assumed DM profile and has much smaller backgrounds. Ideally the best sensitivity to DM properties will be found if the radio measurement is done in a radio ``cold spot", where no known radio sources exist. On the other hand, due to smaller DM density, the synchrotron flux falls away from the GC. This disadvantage is partially mitigated by the excellent sensitivity of the present and upcoming radio telescopes like LWA, LOFAR and SKA.

Finally, anticipating future radio measurements near the GC, we also estimate the constraints that can be obtained on DM self annihilation channels that produce a gamma-ray line using the projected sensitivity of LWA. We very conservatively assume that LWA can reach a background subtracted sensitivity of 10\,Jy at 80\,MHz in a circular region of radius 2$^\circ$ around the GC. After the measurement of the radio flux near the GC one has to model the synchrotron flux due to expected astrophysical processes and then use the uncertainty in that measurement to constrain the DM particle properties (as we have done for the 330 MHz band). Although we will only present our calculated synchrotron fluxes  in regions away from the GC, these can also be used to measure the DM properties. We remind the reader that although the GC is generally opaque to frequencies $\lesssim$ 100\,MHz, the absorption is patchy and the patchiness can be used to do the GC radio measurements~\cite{2003ANS...324Q..65K,1985ApJ...299L..13L,1986Natur.322..522K,2003ANS...324...17B}.

\subsection{Theoretical framework}

To calculate the synchrotron flux from DM self annihilation products, in principle, we need to solve the time-independent diffusion equation for the produced electrons and positrons~\cite{Ishiwata:2008qy}
\begin{eqnarray}
K(E)\nabla^2n_e(E,{\rm \bf{r}}) &+& \dfrac{\partial}{\partial E}\left[b(E, {\rm{\bf r}})n_e(E,{\rm \bf{r}}) \right] \nonumber \\
&=&-S(E, \rm{\bf r})\,,
\label{eq:diffusion equation}
\end{eqnarray}
where $n_e(E,{\rm \bf{r}})$ is the electron density spectrum per unit energy interval, $K(E)$ is the diffusion coefficient, $b(E, \rm{\bf r})$ is the energy-loss rate and $S(E, \rm{\bf r})$ is the source injection spectrum of the electrons. Here $E$ denotes the energy of the electron  and {\rm \bf r} denotes the position of the electron. 

For \mbox{ROI-$2^{\circ}$}, it can be shown that we are in a regime where the GeV electrons will travel only $\sim$ 30\,pc \cite{Crocker:2010gy,Wommer:2008we} during their cooling lifetime. Since this length is much smaller than the length associated with ROI-2$^{\circ}$, we conclude that diffusion will have a small impact on our results. For ROI-4$''$, due to the presence of very high magnetic fields near the GC (see Sec.~\ref{sec:magnetic fields}), and consequently high energy loss rates, the diffusion length $\langle l(E) \rangle \sim \sqrt{KE/b}$, is very small and diffusion can be safely neglected. However, for increased precision, one may in future improve our results by performing a more detailed numerical study along the lines of \cite{Mambrini:2012ue}.

These electrons and positrons then lose energy via the synchrotron process, the inverse Compton process, and the bremsstrahlung process. For our purposes, the total energy loss rate is given by
\begin{eqnarray}
b(E, {\rm{\bf r}})=b_{\rm sync}(E, {\rm{\bf r}})+b_{\rm IC}(E, {\rm{\bf r}})+b_{\rm{brem}}(E,{\rm{\bf r}})\,.
\label{eq:energy loss rate}
\end{eqnarray}
Ionization energy loss is important for electrons and positrons only at lower energies than are considered in this work.

Synchrotron energy losses are due to the interaction of the electron and the positron with the Galactic magnetic field. The energy loss rate due to synchrotron process is given by (all formulae in this section are in SI units)~\cite{Ishiwata:2008qy} 
\begin{eqnarray}
\dfrac{dE}{dt}\bigg|_{\rm sync}&=&\dfrac{4}{3}\sigma_T\,c\,U_{\rm mag}({\rm \bf r}) \gamma ^2 \beta ^2\,
\\ &=&3.4 \times 10^{-17}~{\rm{GeV\, s^{-1}}} \left(\dfrac{E}{{\rm GeV}}\right)^2\left(\dfrac{B({\rm \bf r})}{3\,\mu {\rm G}}\right)^2\,,\nonumber
\label{eq:synchrotron energy loss rate}
\end{eqnarray}
 where $\sigma _T=e^4/(6\pi \epsilon_0^2m_e^2c^4)$ is the Thompson scattering cross section~\cite{Nakamura:2010zzi}, $U_{\rm mag}$ is the magnetic energy density, $\gamma=E/m_e$ is the Lorentz factor, and $\beta=\sqrt{\gamma^2-1}/\gamma$. 
 The photons emitted because of synchrotron energy loss is generally in the radio band.
 
Inverse Compton energy losses are caused by the up-scattering of the photons in the  GC region (which is mainly composed of the CMB and the background starlight) by the more energetic electrons and the positrons. The energy loss rate due to the inverse Compton process is given by~\cite{Ishiwata:2008qy}
\begin{eqnarray}
\dfrac{dE}{dt}\bigg|_{\rm IC}&=&\dfrac{2}{9} \dfrac{e^4 \,U_{\rm rad}({\rm {\bf r}})\,E^2}{\pi \, \epsilon_0^2\, m_e^2 \, c^7} \\
&=&10^{-16}~{{\rm GeV \,s^{-1}}} \left(\dfrac{E}{\,{\rm GeV}}\right)^2\left(\dfrac{U_{\rm {rad}}({\rm \bf r})}{{\rm{eV\, cm^{-3}}}}\right)\,,\nonumber
\label{eq:IC energy loss rate}
\end{eqnarray}
where $U_{\rm rad}$ is the radiation energy density.
The photons from the CMB and the background starlight is generally up scattered to gamma-ray energies by the energetic electrons and positrons from the decays of the $Z$-boson and the Higgs boson.

Bremsstrahlung losses are caused by the emission of gamma-ray photons by the electrons and positrons due to their interaction with the nuclei in the Galaxy. The energy loss rate for this process is given by the Bethe-Heitler formula~\cite{2010hea..book.....L}. We assume that the hydrogen nuclei are the dominant nuclei present in the Galaxy. The energy loss rate due to bremsstrahlung is given by 
\begin{eqnarray}
\dfrac{dE}{dt}\bigg|_{\rm brem}=3\times 10^{-15}~{{\rm GeV \,s^{-1}} } \left(\dfrac{E}{\,{\rm GeV}}\right)\left(\dfrac{n_{\rm {H}}}{{\rm{3\, cm^{-3}}}}\right)\,,\phantom{ 1}
\label{eq:brem energy loss rate in terms of scaled values}
\end{eqnarray}
where we use $n_{\rm {H}}\approx3\,{\rm cm^{-3}}$ as the number density of hydrogen nuclei in the interstellar matter in the Galaxy.

The source term is due to the particle injection from DM self annihilation
\begin{eqnarray}
S(E, {\rm {\bf r}})=\dfrac{1}{2}\langle \sigma v \rangle \left(\dfrac{\rho_{\chi}(\rm {\bf r})}{m_{\chi}}\right)^2 \dfrac{dN_e}{dE}\, ,
\label{eq:source term in diffusion equation}
\end{eqnarray}
where $m_{\chi}$ denotes the mass of the DM particle, $\rho_{\chi}$({\bf r}) denotes the DM density distribution. $dN_e/dE$ denotes the number of electrons and positrons from the decay of the $Z$ or the $H$ boson per unit energy interval, which we calculate using PYTHIA~\cite{Sjostrand:2007gs}. 

Collecting the above mentioned inputs and in the no diffusion limit~\cite{Crocker:2010gy}, we can write electron density spectrum per unit energy interval as
\begin{eqnarray}
n_e(E, {\rm{\bf r}})=-\dfrac{\int_{E}^{m_{\chi}}dE'~S(E', {\rm{\bf r}})}{b(E,{\rm{\bf r}})}\, .
\label{eq:electron energy density spectrum per unit energy interval}
\end{eqnarray}

\subsubsection{Flux density in ROI-2$^\circ$}

The synchrotron power density per unit frequency from a spectrum of electrons and positrons is given by~\cite{1986rpa..book.....R, 2010hea..book.....L}
\begin{eqnarray}
L_{\nu}&({\rm{\bf r}})&=\int dE~ n_e(E, {\rm{\bf r}})\nonumber\\ 
&\times& \left\{\dfrac{1}{4\pi \epsilon _0} \dfrac{\sqrt{3}e^3B({\rm{\bf r}})}{m_e c}
 \left(\dfrac{\nu}{\nu_c}\int _{\nu/\nu_c}^{\infty}dx~K_{5/3}(x) \right) \right\}\, ,
\label{eq:synchrotron power density per unit frequency}
\end{eqnarray}
where the critical frequency, $\nu_c$, is given by 
\begin{eqnarray}
\nu_c=\dfrac{3eE^2B}{4\pi m_e^3 c^4}=16\,{\rm MHz} \left( \dfrac{E}{\rm GeV}\right)^2 \left( \dfrac{B}{\mu{\rm G}}\right)\, ,
\label{eq:critical frequency}
\end{eqnarray}
and $K_{5/3}(x)$ denotes the modified Bessel function of order 5/3.

The synchrotron radiation flux density is given by
\begin{eqnarray}
F_{\nu}=\dfrac{1}{4\pi}\int d\Omega~\int dl~L_{\nu}({\rm{\bf r}})\, ,
\label{eq:synchrotron flux density}
\end{eqnarray}
where $l$ denotes the line of sight distance and $\Omega$ is the angular area of the region of interest.

We have verified that the synchrotron self-absorption is unimportant for these parameters which is generally a problem near the GC at frequencies below approximately 100 MHz but the absorption regions are patchy~\cite{2003ANS...324Q..65K,1985ApJ...299L..13L,1986Natur.322..522K,2003ANS...324...17B}.

\subsubsection{Flux density in ROI-4$''$}

For the smaller region of interest of radius 4$''$ around the GC, i.e., ROI-4$''$, we follow the method presented in Ref.\,\cite{Bertone:2008xr}, which is dependent on the morphology of the magnetic field near the GC black hole. Due to the strong magnetic fields in this region, we assume that the energy loss of the electron is dominated by the synchrotron energy losses. In this case, we approximate
\begin{eqnarray}
\dfrac{\nu}{\nu_c}\int _{\nu/\nu_c}^{\infty}dx~K_{5/3}(x) \approx \dfrac{8\pi}{9\sqrt{3}}~\delta\left(\dfrac{\nu}{\nu_c}-\dfrac{1}{3} \right)\, .
\label{eq:delta function for integral of the Bessel K function}
\end{eqnarray}

The synchrotron flux density in this case is given by~\cite{Bertone:2008xr}
\begin{eqnarray}
F_{\nu}=\dfrac{1}{4\pi({\rm{8.5~kpc}})^2} \dfrac{\langle \sigma v \rangle}{2 m_{\chi}^2}  
\int dV \rho_{\chi}^2 E   \int _{E}^{m_{\chi}} \dfrac{dN}{dE'} dE'\, ,\phantom{ 1}
\label{eq: synchrotron flux density delta function approximation}
\end{eqnarray}
where the first integral is over the volume of observation and the second integral counts the number of particles above a certain energy $E$. The value of $E$ in this case can be found by using Eqs.\,(\ref{eq:critical frequency}) and\,(\ref{eq:delta function for integral of the Bessel K function}) and is given by
\begin{eqnarray}
E=433\, {\rm MeV} \sqrt{\left(\dfrac{\nu}{\rm MHz} \right)\left(\dfrac{\mu {\rm G}}{B} \right)}\,.
\label{eq:energy in ROI - 4"}
\end{eqnarray}

\subsubsection{Flux density in ROI-away}
 We calculate the synchrotron flux in $ROI$-$away$ in the same way as we do in $ROI$-$2^\circ$. To account for the fact that we are now calculating the synchrotron flux away from the GC, we do make some modifications to our input of the interstellar proton density and the interstellar radiation density. For simplicity, we take the number density of hydrogen nuclei in the interstellar medium of our Galaxy to be 3\,cm$^{-3}$. We take the variation of the radiation energy density following Ref.\,\cite{Ishiwata:2008qy}. More details about the variation of the radiation energy density is given in Sec.~\ref{sec:radiation}.

\section{Astrophysical inputs for calculations}
\label{sec:Inputs}
\subsection{DM density profiles}

One of the major unknowns near the GC is the DM density profile. Almost all simulations agree on the radial dependence of the DM density profile  at large radii from the GC, $\rho(r)$ $\sim$ $r^{-3}$. However, due to limited numerical resolution and the complicated astrophysics at the GC, the simulations disagree on the shape of the density profile at small radii. 

Observations of elliptical galaxies~\cite{2010MNRAS.408.1463S}, early-type galaxies~\cite{2010ApJ...721L.163A}, M31~\cite{2008MNRAS.389.1911S}, and M84~\cite{2011MNRAS.411.2035N} prefer a cuspy profile~\cite{Gnedin:2011uj} in contrast to dwarf galaxies which prefer a cored profile~\cite{deBlok:2009sp}. Given that M31 is a Milky Way like galaxy, we assume that the DM density profile in the Milky Way is not cored and hence we do not consider the cored isothermal profile in our work. In general, the constraints from the indirect detection searches are especially weak for a cored isothermal profile~\cite{Crocker:2010gy, Bertone:2008xr}.

\begin{figure*}[!t]
\centering
\includegraphics[angle=0.0, width=0.31\textwidth]{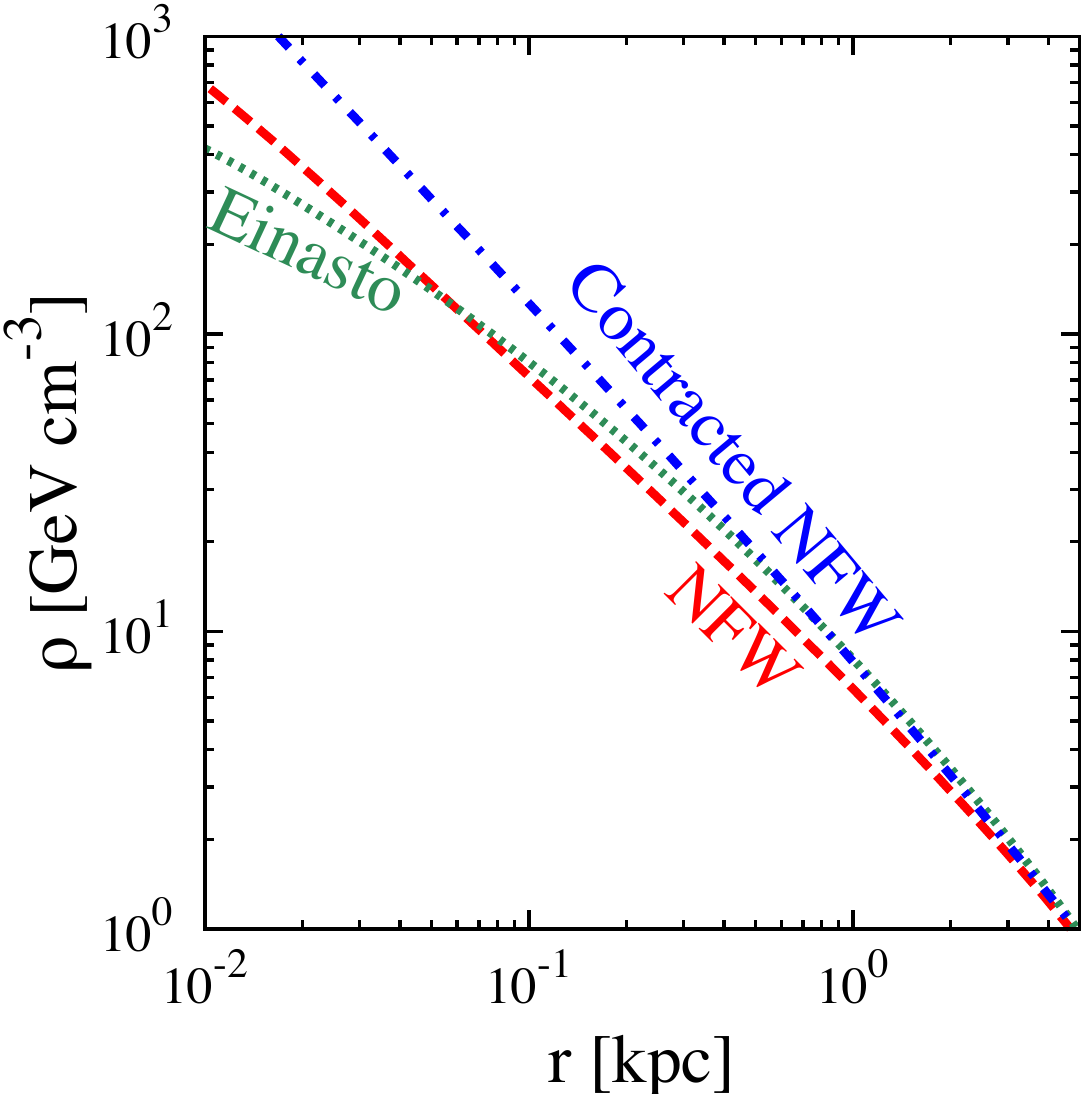}
\includegraphics[angle=0.0, width=0.323\textwidth]{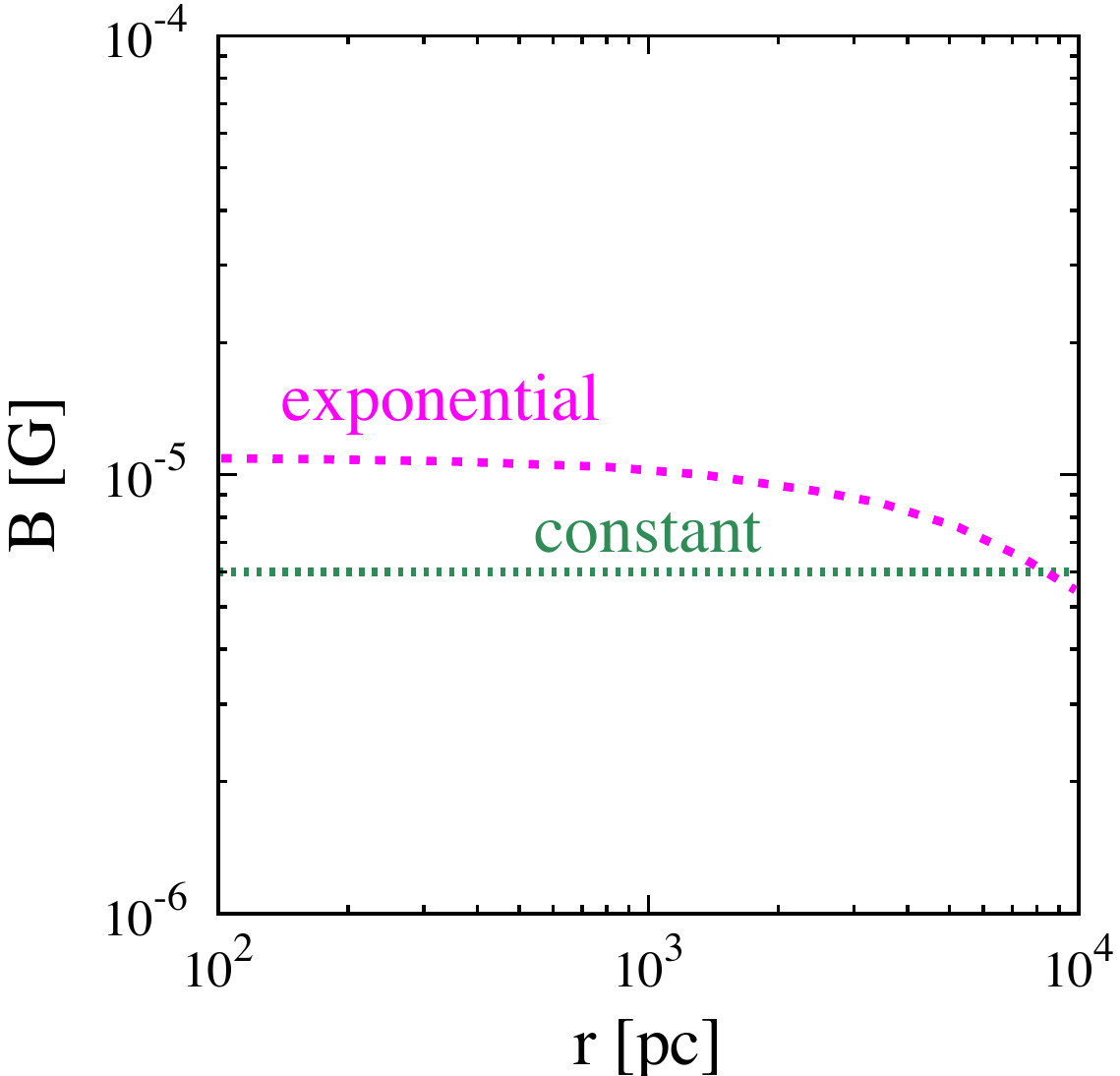}
\includegraphics[angle=0.0, width=0.323\textwidth]{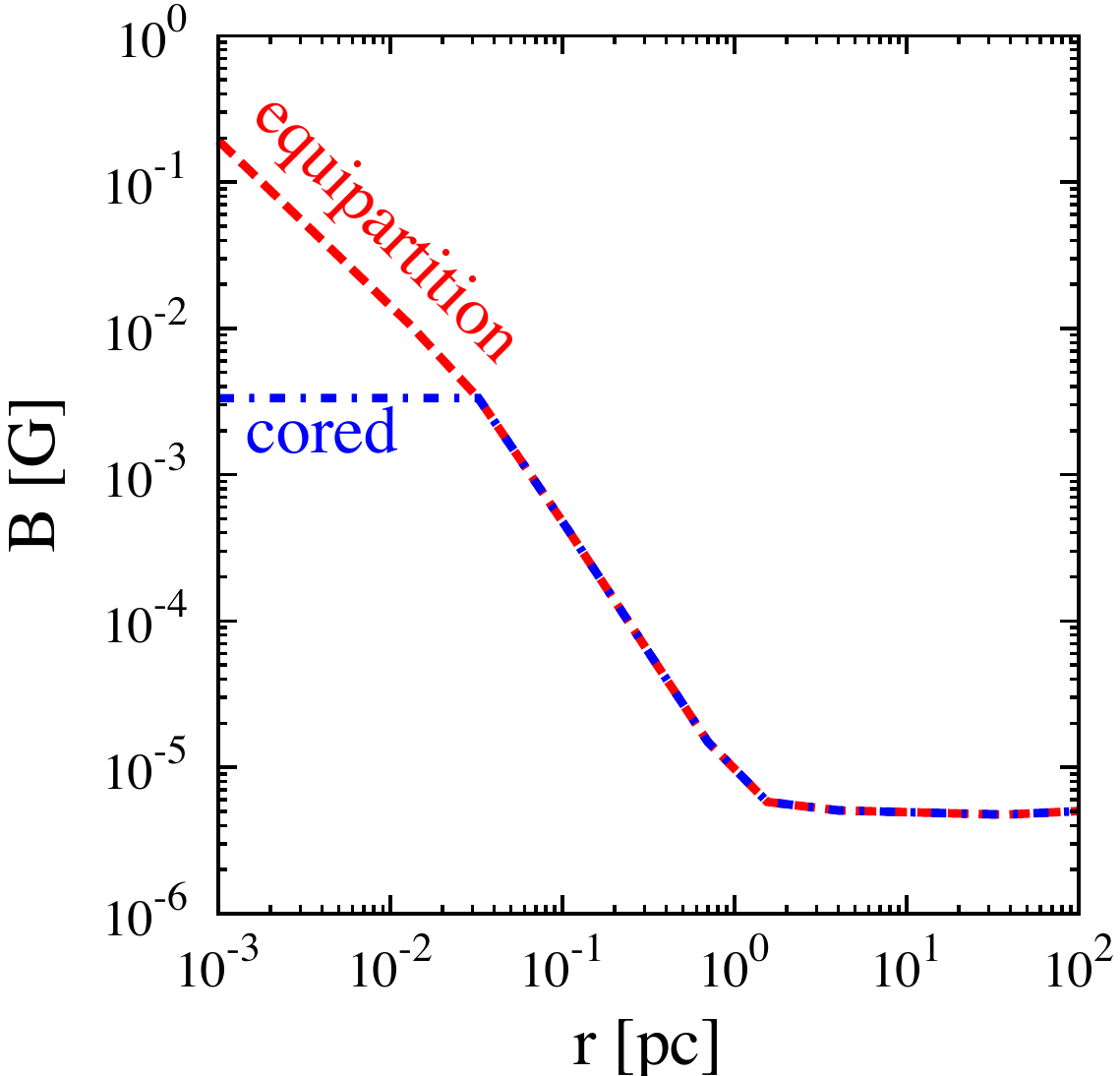}
\caption{Galactic dark matter density {(left panel)} and magnetic field profiles {(middle and right panels)}. {\bf (Left)} We show the various DM profiles used in this work: the Einasto profile (\ref{eq:Einasto}), the NFW profile (\ref{eq:NFW}) and the contracted NFW profile (\ref{eq:contracted NFW}). {\bf (Middle)} We show the magnetic field used for calculating the synchrotron flux in ROI-2$^\circ$, i.e., a region of angular radius 2$^\circ$ around the GC. The constant magnetic field has a value of 6\,$\mu$G everywhere in the Galaxy. The exponential magnetic field is given in Eq.\,(\ref{eq:large scale galactic magnetic field}) and has a value of 6\,$\mu$G at the solar radius. {\bf (Right)} We show the magnetic field used for calculating the synchrotron flux in ROI-4$''$, i.e., in a region of angular radius 4$''$ around the GC. The ``equipartition" magnetic field is given by Eq.\,(\ref{eq:equipartition magnetic field}) and the ``cored" magnetic field is given by Eq.\,(\ref{eq:cored magnetic field}). Both the fields have been normalized to have a value of 6\,$\mu$G at the solar radius.}
\label{fig:DM & magnetic field}
\end{figure*}

In this work, we use three different DM profiles which provide reasonable constraints from the radio measurements at the GC. The dark matter density at the solar radius has a value of 0.3 $\pm$ 0.1~GeV/cm$^{3}$~\cite{Bovy:2012tw}. For concreteness, in this work, we take the benchmark value to be 0.4 GeV/cm$^3$~\cite{Catena:2009mf,Salucci:2010qr}. Note that local dark matter density taken in the papers which discuss the presence of the 130 GeV line at the GC is also 0.4\,GeV/cm$^3$ (see, e.g., Refs.~\cite{Weniger:2012tx,Tempel:2012ey}).

The Einasto dark matter profile~\cite{1965TrAlm...5...87E, Navarro:2003ew, Springel:2008cc, Pieri:2009je},
\begin{eqnarray}
\rho_{\rm{Ein}}(r)&=&\dfrac{0.08~{{\rm{GeV}\,cm}^{-3}}}
{{\rm{exp}}\left[\dfrac{2}{0.17}\left(\left(\dfrac{r}{20~{\rm{kpc}}}\right)^{0.17}-1\right)\right]^{\phantom{ 1}}}\, .
\label{eq:Einasto}
\end{eqnarray}
is the least cuspy of all the DM profiles considered in this work, and hence we expect this profile to produce least amount of synchrotron radiation, especially when we consider the synchrotron radiation from a region very near the GC.

We then consider the standard NFW profile~\cite{Navarro:1996gj}
\begin{eqnarray}
\rho_{\rm{NFW}}(r)=\dfrac{0.35~{\rm{GeV}\,cm}^{-3}}{\left(\dfrac{r}{20~{\rm{kpc}}} \right)\left(1+\dfrac{r}{20~{\rm{kpc}}} \right)^{2{^{\phantom{1}}}}}\, .
\label{eq:NFW}
\end{eqnarray}
The cuspy nature of this DM density profile will ensure that we get a larger synchrotron radiation flux than what we expect from the Einasto profile when we consider observation from a region very near to the GC.

We finally consider a contracted NFW profile
\begin{eqnarray}
\rho_{\rm{con.\,NFW}}(r)=\dfrac{0.29~{\rm{GeV}\,cm}^{-3}~}{\left(\dfrac{r}{20~{\rm{kpc}}} \right)^{1.15}\left(1+\dfrac{r}{20~{\rm{kpc}}} \right)^{1.85}}.
\label{eq:contracted NFW}
\end{eqnarray}
The steeper inner slope in this case can be due to either a GC black hole~\cite{Gondolo:1999ef}, or due to adiabatic contraction due to the presence of baryonic matter at the GC~\cite{Blumenthal:1985qy, 1987ApJ...318...15R, Gnedin:2003rj, Gustafsson:2006gr}, which have been supported by more recent numerical simulations~\cite{Gnedin:2011uj}. 

These DM profiles are shown in the left panel in Fig.\,\ref{fig:DM & magnetic field}. It is evident from the figure that, at small radii, the contracted NFW profile has the steepest slope, and the Einasto profile has the shallowest slope of all the three DM profiles considered in this work. From the figure, one can also infer that the DM density profiles have almost the same shape at large distances from the GC.

\subsection{Magnetic fields}
\label{sec:magnetic fields}

The GC magnetic field has both a regular and a turbulent component. For both the components, the normalization and the radial profile is not understood very well. In particular, the magnetic field amplitude near the GC is uncertain, with measured estimates spanning a range of some two orders of magnitude between $10 \, {\rm \mu G}$~\cite{LaRosa:2005ai} and $10^3 \, {\rm \mu G}$ on scales of a few \mbox{$\sim\,100$\,pc~\cite{1989ApJ...343..703M}}. In order to account for the uncertainty in the magnetic field, we adopt several configurations. In all cases, we initially fix the normalization to a value $B_\odot = 6 \, \mu$G at the solar system radius ($r_\odot = {\rm 8.5 \, kpc}$). This is mid-range among the various estimates of $B_\odot$ which span between 3\,$\mu$G and 10\,$\mu$G~\cite{Han:2006ci, Jansson:2009ip, Pshirkov:2011um}. We will discuss how our results scale with the different values of $B_\odot$ when we present our results. 

\subsubsection{Magnetic fields in ROI-2$^\circ$}
For ROI-2$^\circ$, i.e., a circular region with radius 2$^\circ$ around the GC (distance scale $\sim 200$\,pc for $r_\odot= 8.5$ kpc), we consider two different magnetic field radial profiles. The first is the spherically symmetric exponential profile,
\begin{eqnarray}
B(r)=B_\odot ~{\rm exp}\left(-\frac{r- r_\odot}{R_m} \right)\, ,
\label{eq:large scale galactic magnetic field}
\end{eqnarray}
where $r$ is the distance from the GC, and $R_m = 14 \, {\rm kpc}$ is the scale radius. Our choice of $R_m$ follows from Ref.\,\cite{Fornengo:2011iq}, where we adopt their Galactic magnetic field model ``GMF I'' and their best-fit propagation parameters. We add that we neglect the $z$-dependence of the magnetic field which is only weakly constrained by data and remain highly uncertain. Using this conservative form of the magnetic field, the magnetic field at a radius of 2$^\circ$ is $\approx 11 \, {\rm \mu G}$. Although this value is within the range of estimates of the magnetic field in the GC, it is closer to the lower range. In addition, it does not obey the lower limit of $50 \, {\rm \mu G}$ on scales of $400$\,pc presented in Ref.\,\cite{2010Natur.463...65C}. However, given the uncertainties in the astrophysical and propagation quantities, we do not consider this difference significant. For example, if we adopt instead $B_\odot = 10 \, \mu$G and the ``MAX'' propagation parameters of Ref.\,\cite{Fornengo:2011iq}, we obtain $R_m \approx 8.5 \, {\rm kpc}$ and a magnetic field at 2$^\circ$ of $\approx 27 \, {\rm \mu G}$. To estimate the impact of the normalization of the magnetic field, we will also show our results for two extreme values for $B_{\odot}$, i.e., 3\,$\mu$G and 10\,$\mu$G. 

To estimate the uncertainty due to the shape of the magnetic field profile we also adopt the extreme case of a constant magnetic field of value $B_\odot$ everywhere~\cite{Hooper:2007kb}. Both of these are shown on the middle panel of Figure \ref{fig:DM & magnetic field}. 

\subsubsection{Magnetic fields in ROI-4$''$}
For ROI-4$''$, i.e., a region with radius 4$''$ around the GC (distance scale $\sim 0.2$\,pc for $r_\odot= 8.5$ kpc), we need to take into account the influence of the supermassive black hole (SMBH) at the GC. The presence of the SMBH sets two length-scales: the Schwarzschild radius, $R_{\rm BH} \approx 1.2 \times 10^{12} (M_{\rm}/ 4.3 \times 10^6 M_\odot)$\,cm, and the radius $R_{\rm acc} \approx 0.04$\,pc within which the free-fall velocity due to the gravity of the SMBH $v = -c \sqrt{R_{\rm BH}/ r}$ is larger than the random Galactic motion, $\sim 10^{-3} c$. In other words, the region $r < R_{\rm acc}$ defines the accretion region. 

We adopt the ``equipartition model'' for the Galactic magnetic field, described by various authors \cite[e.g.,][]{Bertone:2008xr}. In this model, the SMBH accretes matter from a radius of $ R_{\rm acc}$, and the magnetic field in the accretion flow achieves its equipartition with the kinetic pressure, i.e., $B^2(r) / (2\mu_0) = \rho(r) v^2(r) /2$. For a constant mass accretion rate, $\dot{M}$, one obtains $\rho (r) = \dot{M} / 4 \pi r^2 v(r) \propto r^{-3/2}$, and thus $B(r) \propto r^{-5/4}$. Outside of $R_{\rm acc}$, the conservation of magnetic flux is assumed, yielding $B(r) \propto r^{-2}$. Thus, the equipartition magnetic field is given by
\begin{eqnarray} 
\label{eq:equipartition magnetic field}
B_{\rm eq}(r) = 
\left\{
\begin{array}{lll}
B_{\rm acc} (r/R_{\rm acc})^{-5/4} 	&  \,\, r \leq R_{\rm acc}  \\
B_{\rm acc} (r/R_{\rm acc})^{-2} 	&  \,\, R_{\rm acc} < r \leq R_{\rm flux}  \\
B_\odot 						&  \,\, R_{\rm flux} < r~,  
\end{array} \right.
\end{eqnarray}
where $R_{\rm acc}\approx0.04$\,pc, $R_{\rm flux}\sim100\,R_{\rm acc}$, and
\begin{equation}
B_{\rm acc} = 7.9 \, {\rm mG} 
\left( \frac{M_{\rm BH}}{4.3 \times 10^6 M_\odot} \right)^{1/4}
\left( \frac{ \dot{M} }{10^{22} \, {\rm g/s}} \right),
\end{equation}
for typical values of $B_\odot$.

We also consider a variant of the equipartition model, where the inner magnetic field is kept smaller because equipartition is prevented somehow. This may occur if, for example, magnetic field dissipation occurs by reconnection in the turbulent accretion flow (see, e.g., \cite{Bergstrom:2006ny} and references therein). Since the details of dissipation are not well understood, we conservatively adopt a constant magnetic field throughout the accretion region, namely, 
\begin{eqnarray} \label{eq:cored magnetic field}
B_{\rm cored}(r) = 
\left\{
\begin{array}{lll}
B_{\rm acc} 				 	&  \,\, r \leq R_{\rm acc}  \\
B_{\rm acc} (r/R_{\rm acc})^{-2} 	&  \,\, R_{\rm acc} < r \leq R_{\rm flux}  \\
B_\odot 						&  \,\, R_{\rm flux} < r~.
\end{array} \right.
\end{eqnarray}
We call this the cored magnetic field.
These are shown on the right panel of Figure \ref{fig:DM & magnetic field}. 

\subsubsection{Magnetic fields in ROI-away}

While calculating the magnetic field in a region offset from the Galactic Center, we assume the exponential magnetic field structure as in Eq.\,(\ref{eq:large scale galactic magnetic field}).

\subsection{Radiation energy density}
\label{sec:radiation}

The radiation energy density is the sum of the energy density of the CMB photons and the energy density of the background starlight photons. The energy density of the CMB photons is 0.3 eV cm$^{-3}$. The radiation energy density due to the background starlight varies with position  in the Galaxy from 1 eV cm$^{-3}$ to 10~eV~cm$^{-3}$. Conservatively, when we calculate the synchrotron flux in a region near the GC, i.e., in $ROI$-$2^\circ$ and $ROI$-4$"$, we take the background starlight energy density to have a constant value of 9 eV cm$^{-3}$. Hence we use the total radiation field energy density as $U_{\rm{rad}} = 9$~eV~cm$^{-3}$ in this work while calculating the synchrotron flux in a region near the GC. We also check our results by taking a much smaller radiation energy density of 0.9 eV/cm$^3$ and find that using this lower value of the radiation field energy density improves our constraints by a factor of $\sim$ 2 - 3.

When we calculate the synchrotron flux in a region away from the GC, $ROI$-$away$, we follow the radiation energy density parametrization in Ref.\,\cite{Ishiwata:2008qy} which uses the results in Ref.\,\cite{Strong:1998fr}. The energy density at any given position in the Galaxy is~\cite{Ishiwata:2008qy}
\begin{eqnarray}
\label{eq:radiation energy density}
U_{\rm rad}(r,z)=\dfrac{U_{\rm stellar}({\rm 4\,kpc},z)}{U_{\rm stellar}({\rm 4\,kpc},0)}U_{\rm stellar}(r,0)+ U_{\rm CMB}^{\phantom{  1}}\,,\quad
\label{eq:radiation density}
\end{eqnarray}
where we denote the CMB energy density as $U_{\rm CMB}$ and the stellar radiation energy density by U$_{\rm stellar}$. The vertical distance from the plane of the Galaxy is denoted by $z$ and the radial distance from the center of the Galaxy is denoted by $r$. Our choice of the radiation field density is also consistent with~\cite{Moskalenko:2005ng}.

\begin{figure*}[!thbp]
\centering
\includegraphics[angle=0.0, width=0.43\textwidth]{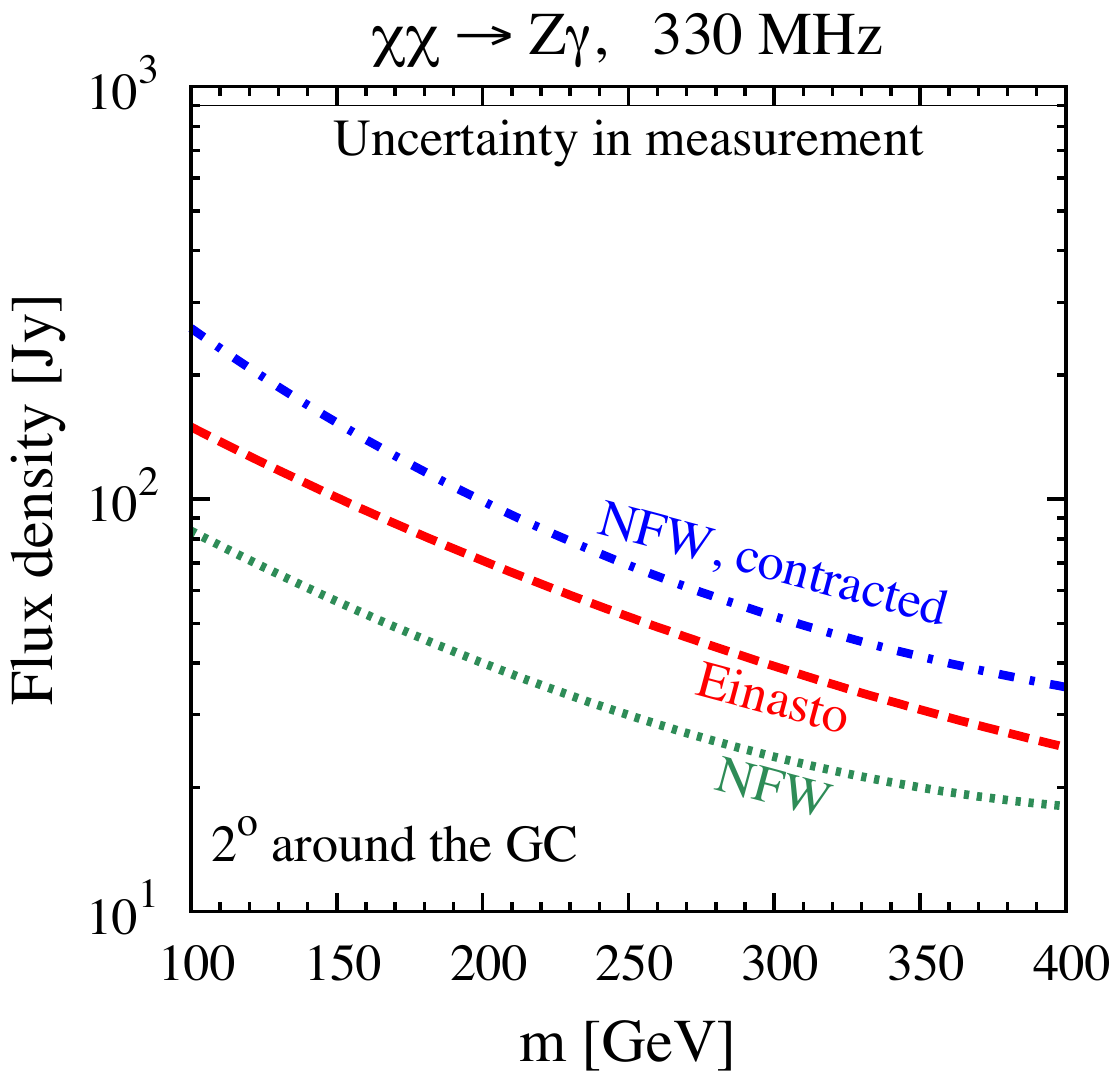}
\includegraphics[angle=0.0, width=0.43\textwidth]{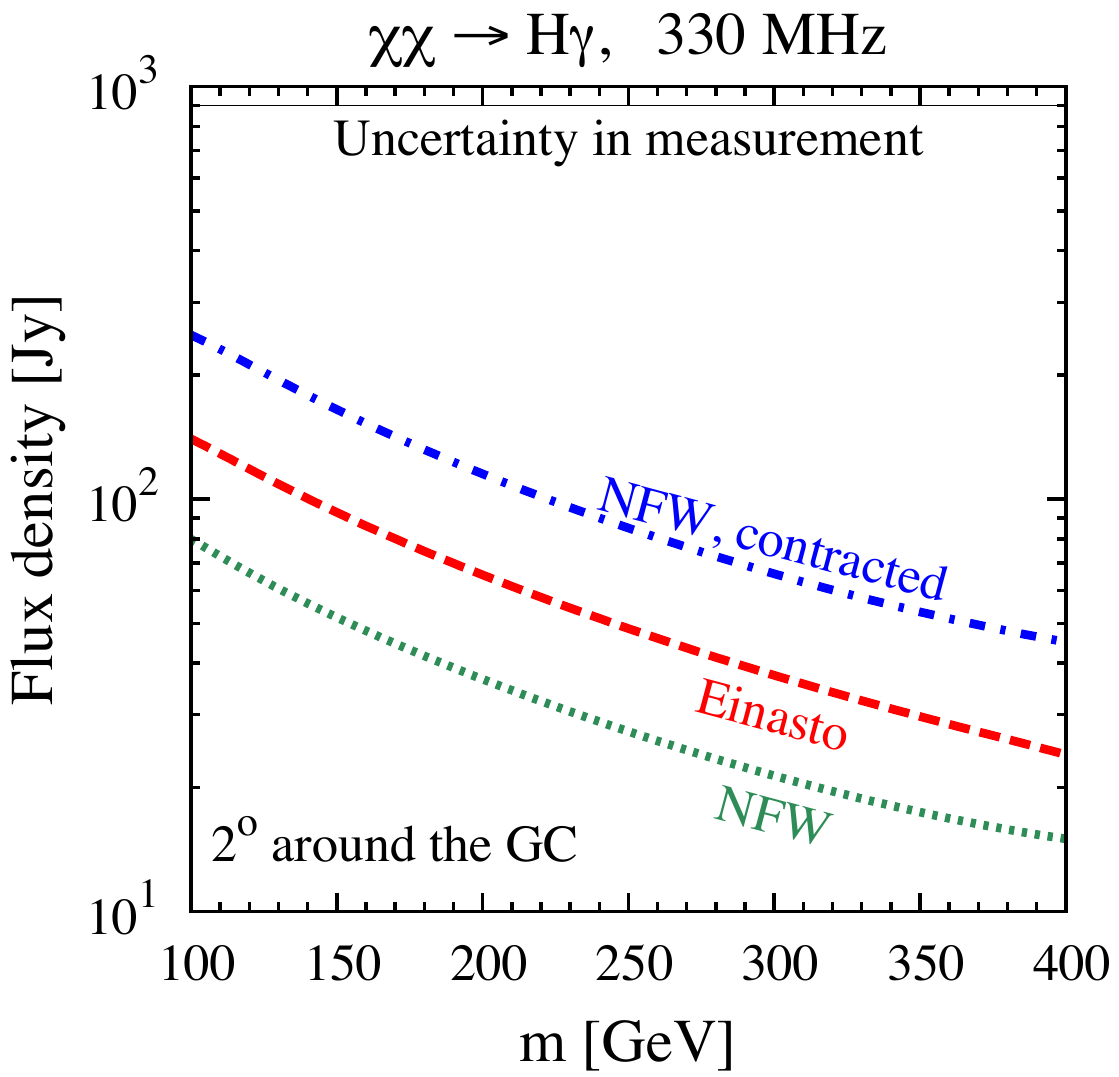}
\includegraphics[angle=0.0,width=0.43\textwidth]
{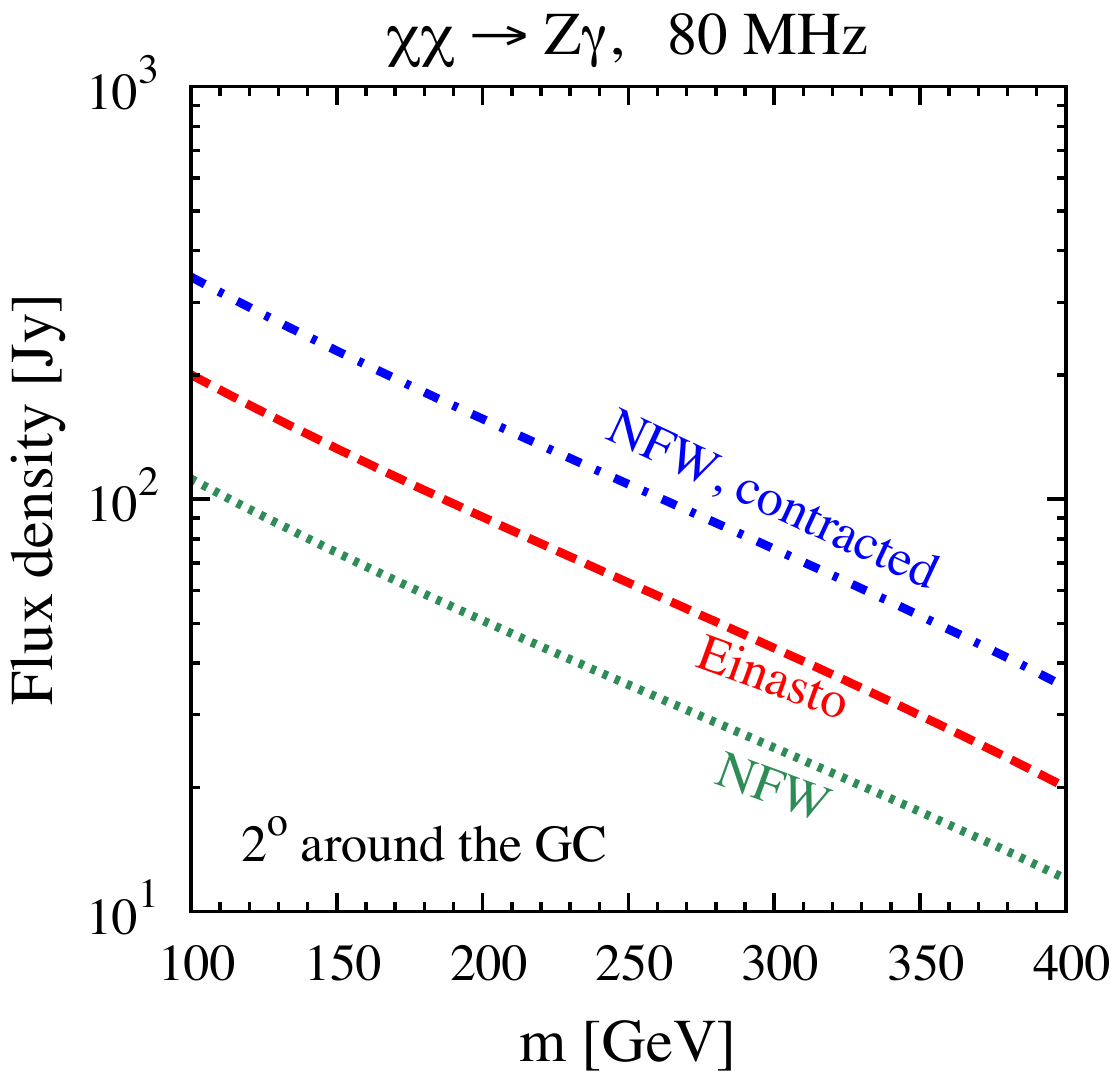}
\includegraphics[angle=0.0,width=0.43\textwidth]
{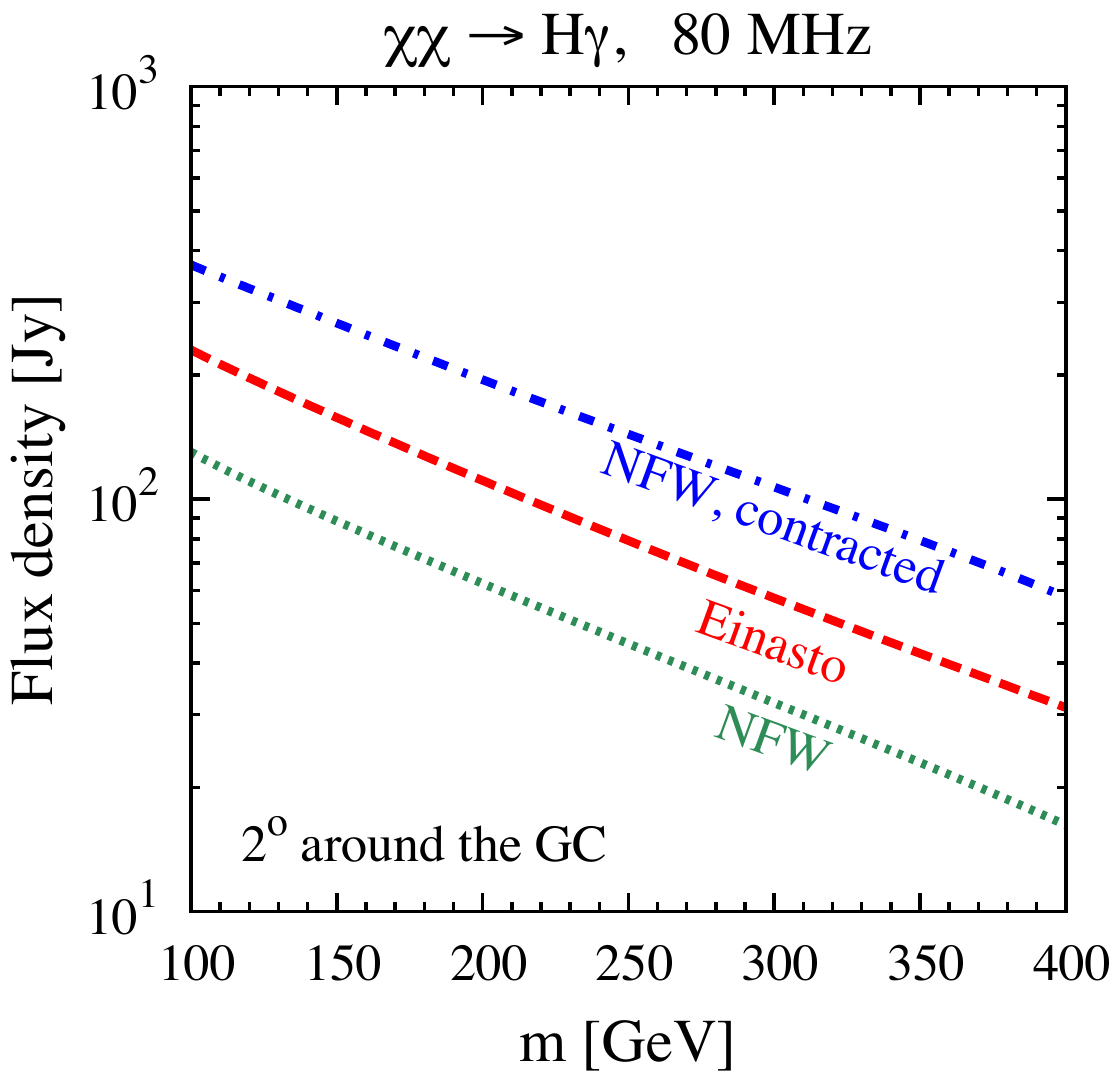}
\caption{Prediction of the synchrotron flux density $2^\circ$ around the GC, against mass of the DM.  In all the plots, we set as a benchmark DM annihilation cross section $\langle \sigma v \rangle=10^{-26}$ cm$^3$ s$^{-1}$, and consider three different DM profiles: Einasto profile in~Eq.\,(\ref{eq:Einasto}), NFW profile in~Eq.\,(\ref{eq:NFW}) and the contracted NFW profile in~Eq.\,(\ref{eq:contracted NFW}). ({\bf Left}) $\chi \chi \rightarrow Z \gamma$. ({\bf Right}) $\chi \chi \rightarrow H \gamma$. ({\bf Top}) Results for 330\,MHz. We also show the uncertainty in the measurement which is used to derive our constraints in this frequency band. ({\bf Bottom}) Results for 80\,MHz. For both the frequency bands, we use the exponential magnetic field in Eq.\,(\ref{eq:large scale galactic magnetic field}).}
\label{fig:flux density ROI2}
\end{figure*}

\begin{figure*}[!thbp]
\centering
\includegraphics[angle=0.0,width=0.43\textwidth]{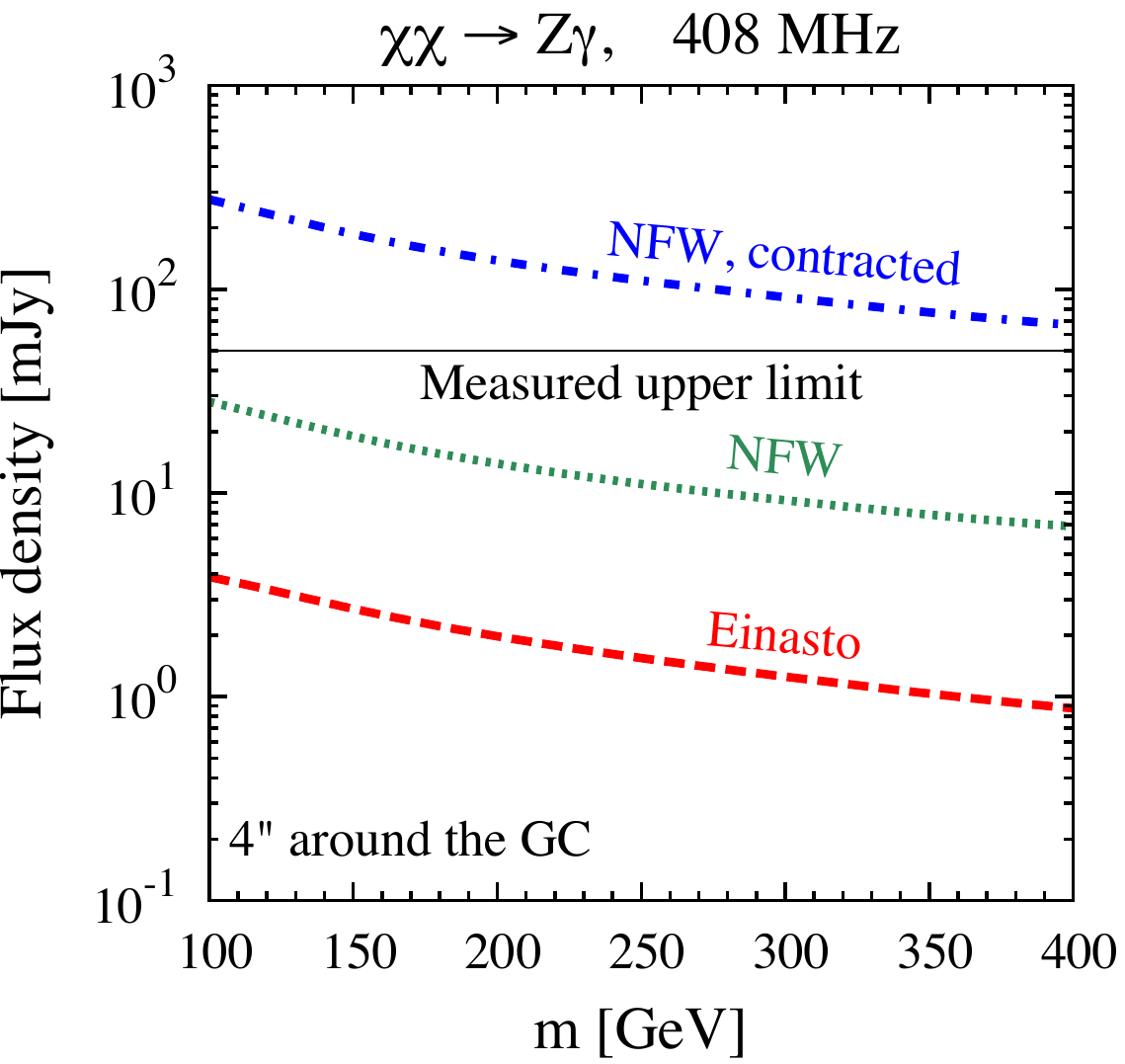}
\includegraphics[angle=0.0,width=0.43\textwidth]{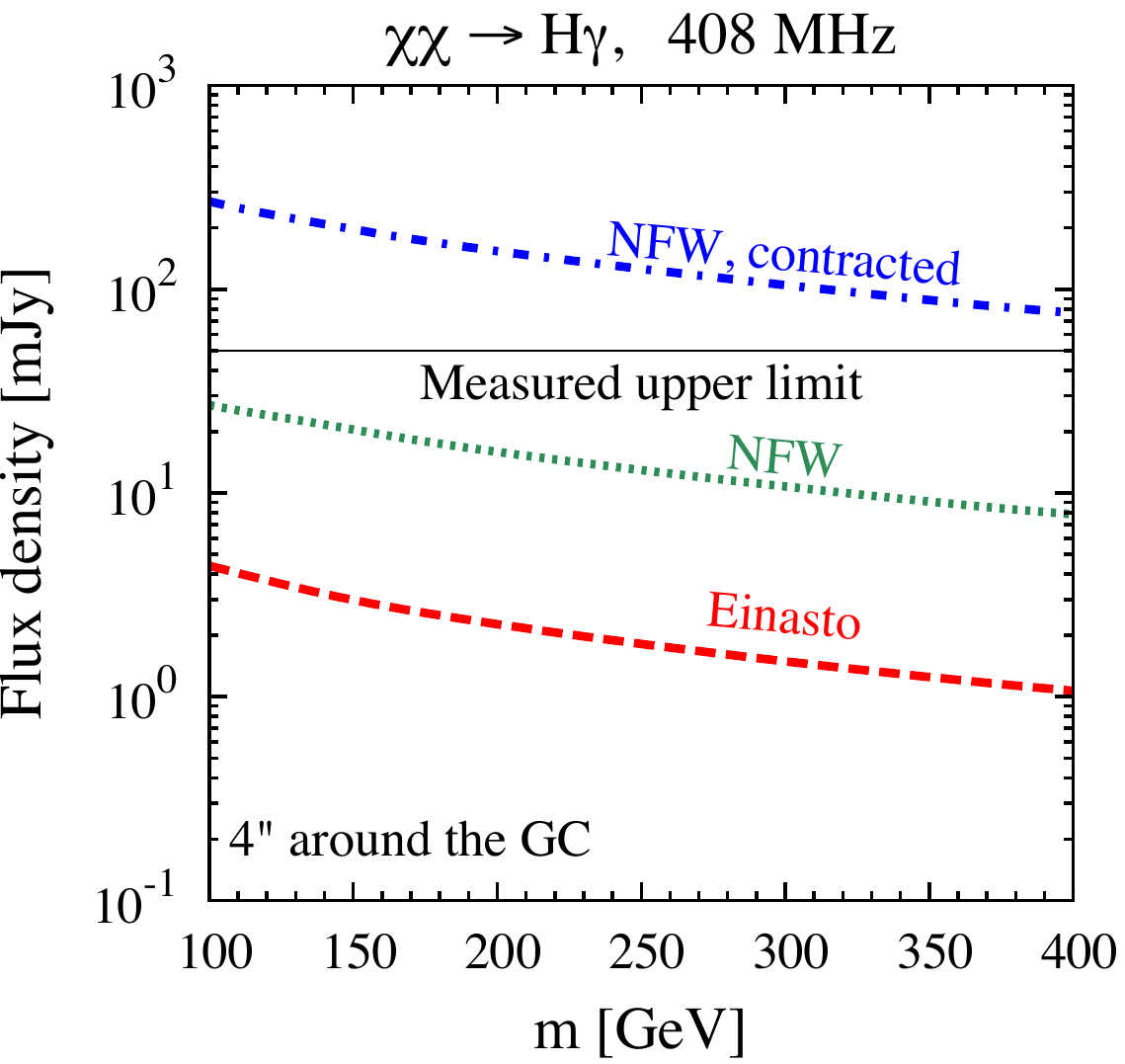}
\includegraphics[angle=0.0,width=0.43\textwidth]{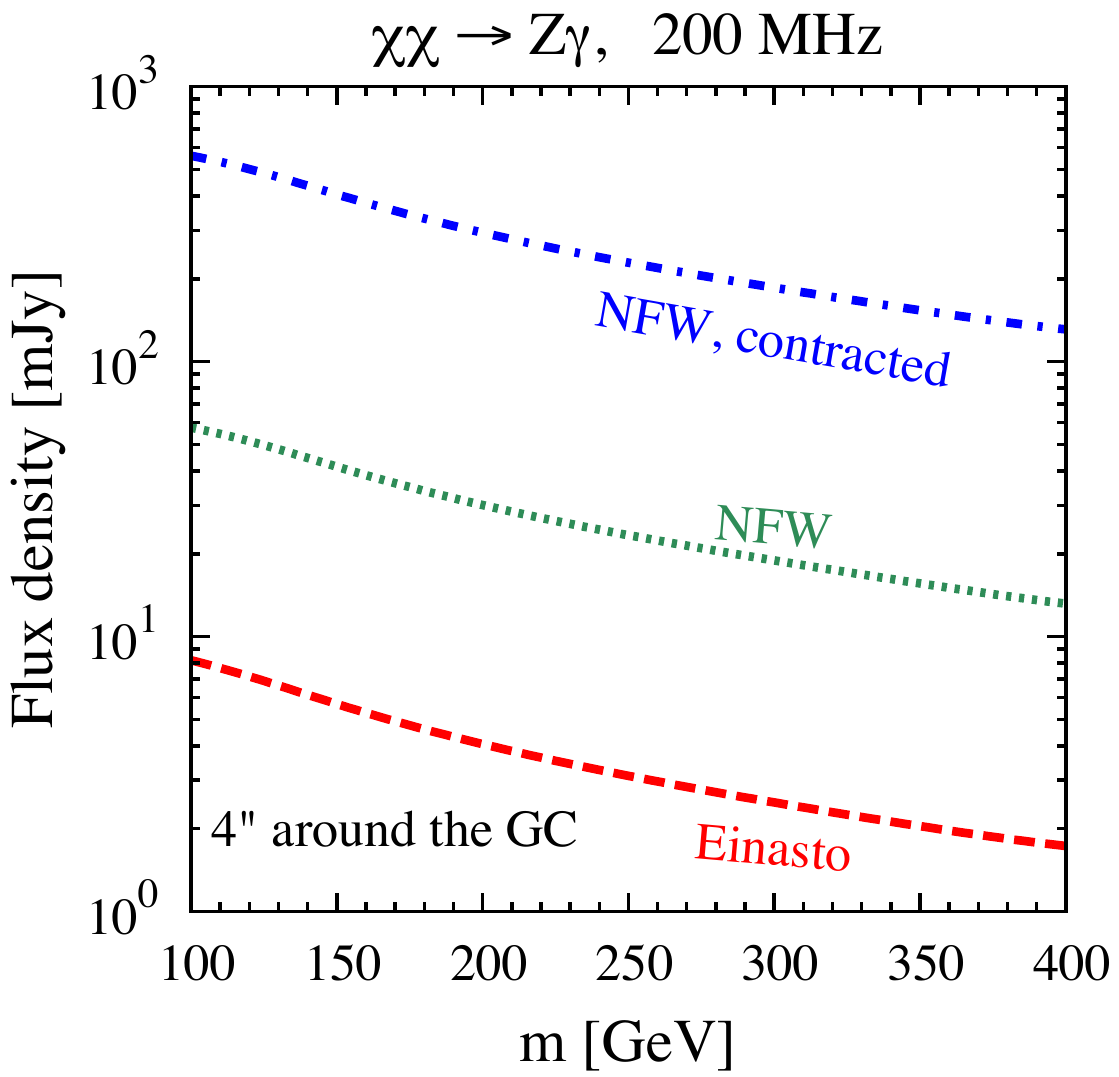}
\includegraphics[angle=0.0,width=0.43\textwidth]{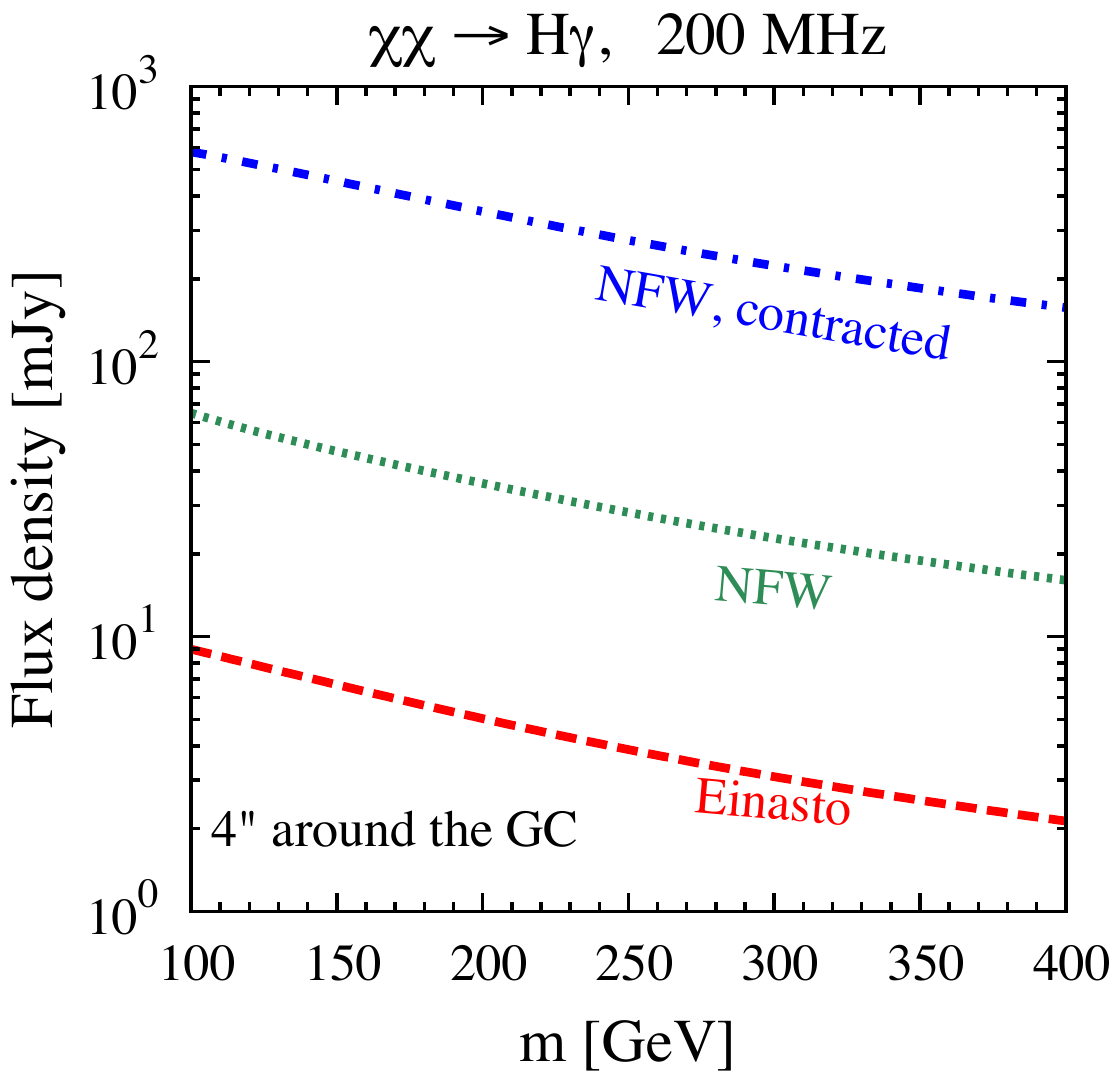}
\caption{Prediction of the synchrotron flux density $4''$ around the GC, against mass of the DM.  The DM annihilation cross section and the profiles are the same as in Fig.~\ref{fig:flux density ROI2}. ({\bf Left}) $\chi \chi \rightarrow Z \gamma$. ({\bf Right}) $\chi \chi \rightarrow H \gamma$. ({\bf Top}) Results for 408\,MHz. ({\bf Bottom}) Results for 200\,MHz. We use the equipartition magnetic fields in Eq.\,(\ref{eq:equipartition magnetic field}) for both these frequency bands.}
\label{fig:flux density ROI4}
\end{figure*}

\begin{figure*}[!thbp]
\centering
\includegraphics[angle=0.0,width=0.43\textwidth]{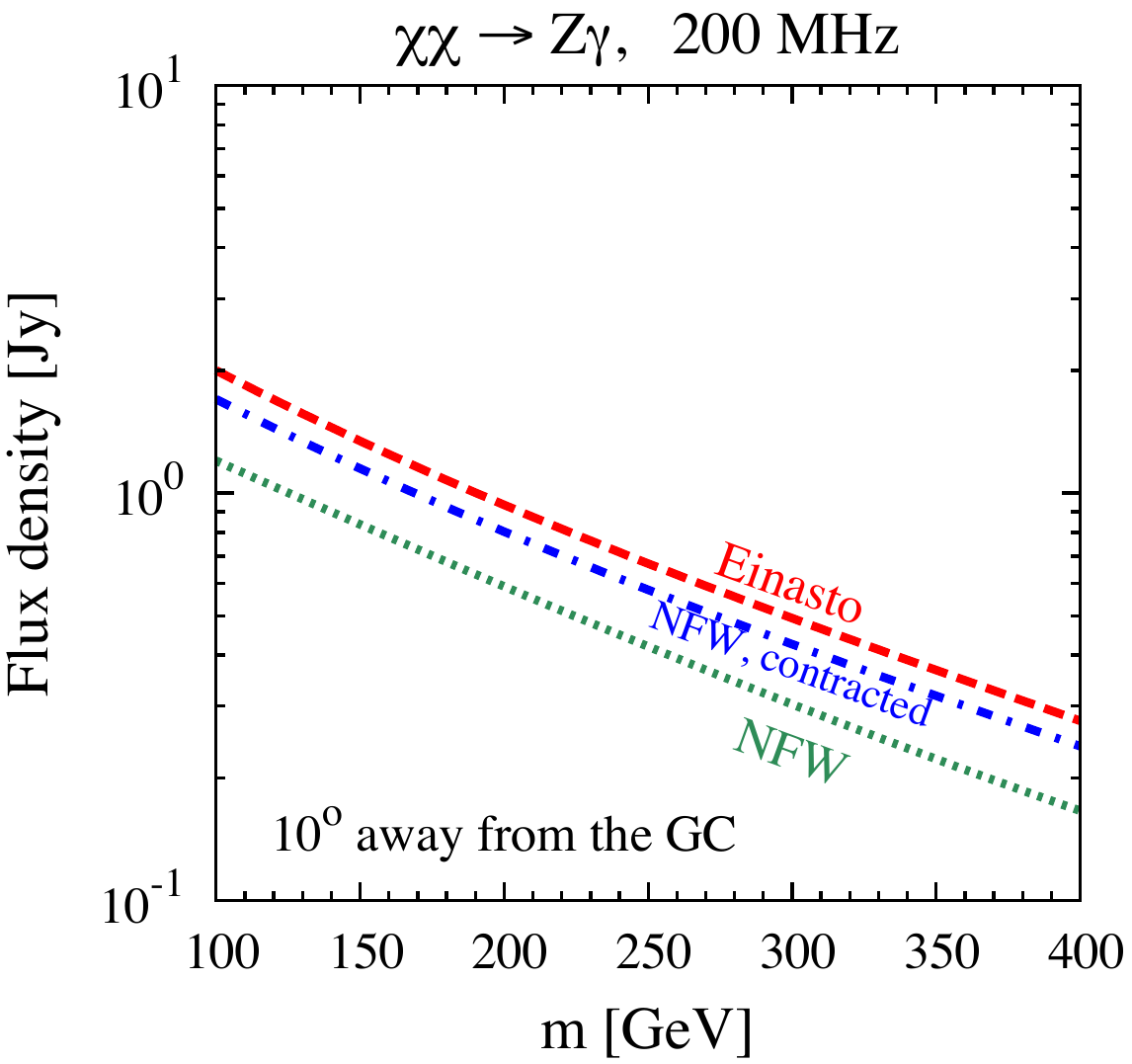}
\includegraphics[angle=0.0,width=0.43\textwidth]{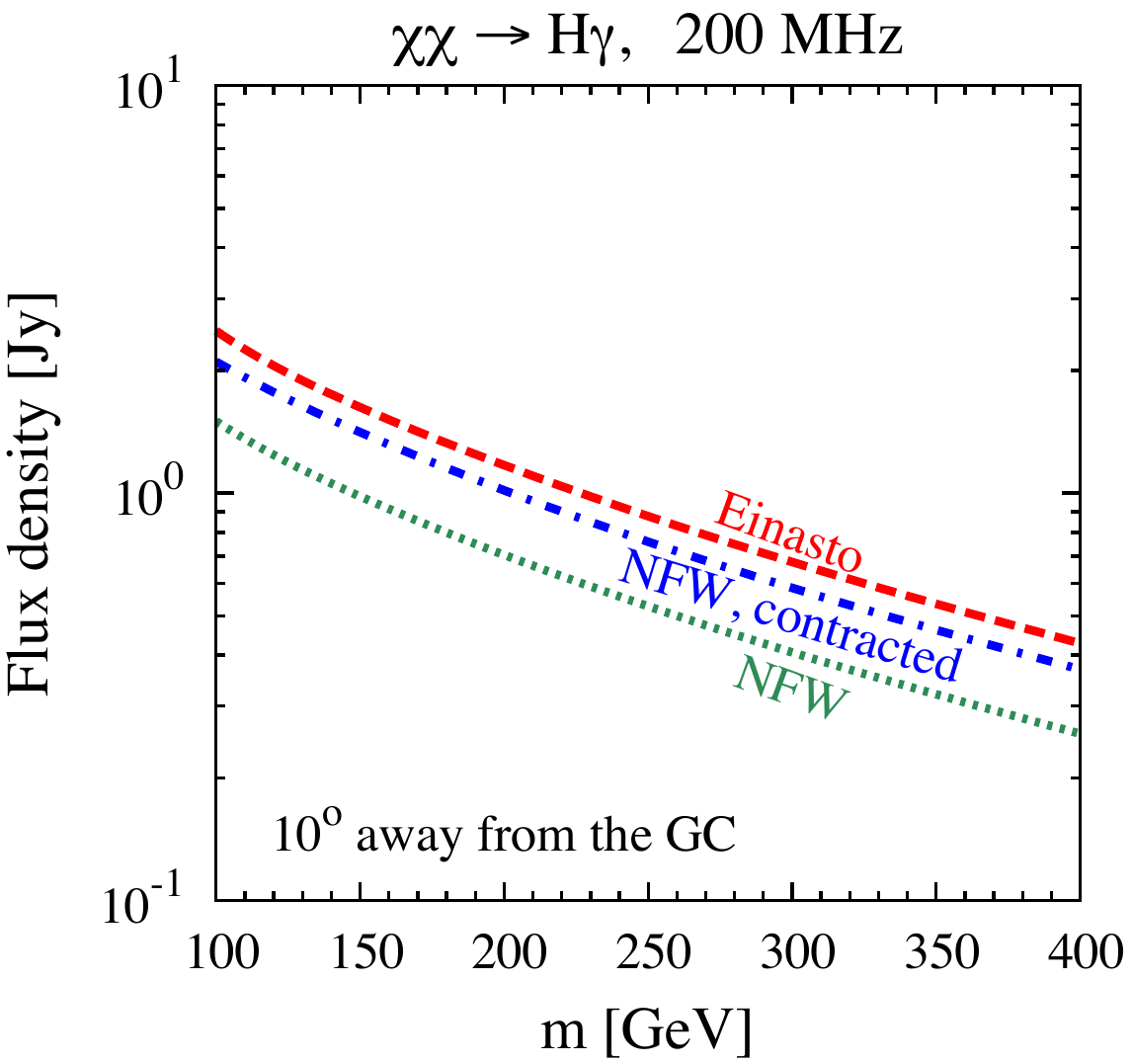}
\includegraphics[angle=0.0,width=0.43\textwidth]
{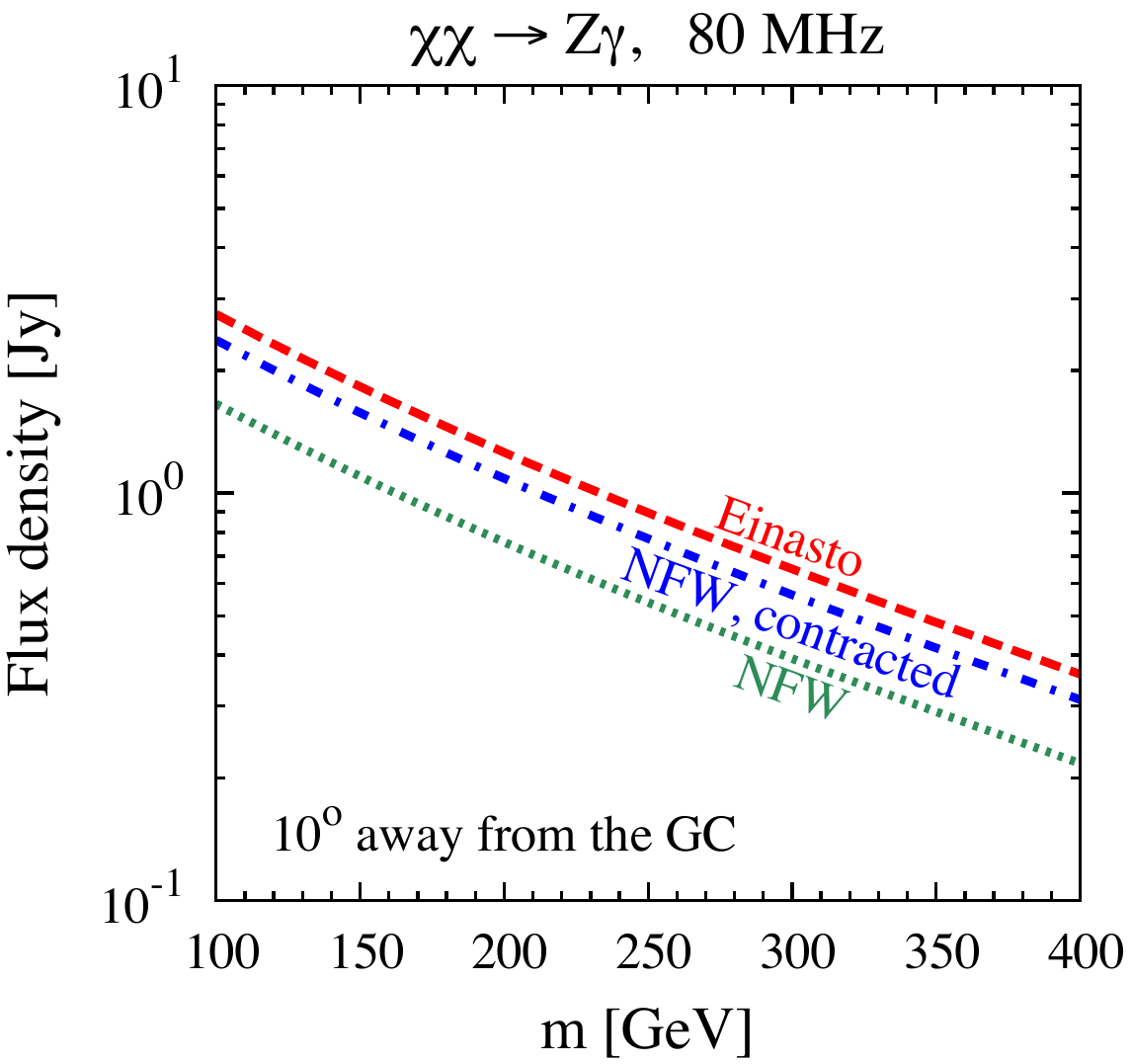}
\includegraphics[angle=0.0,width=0.43\textwidth]{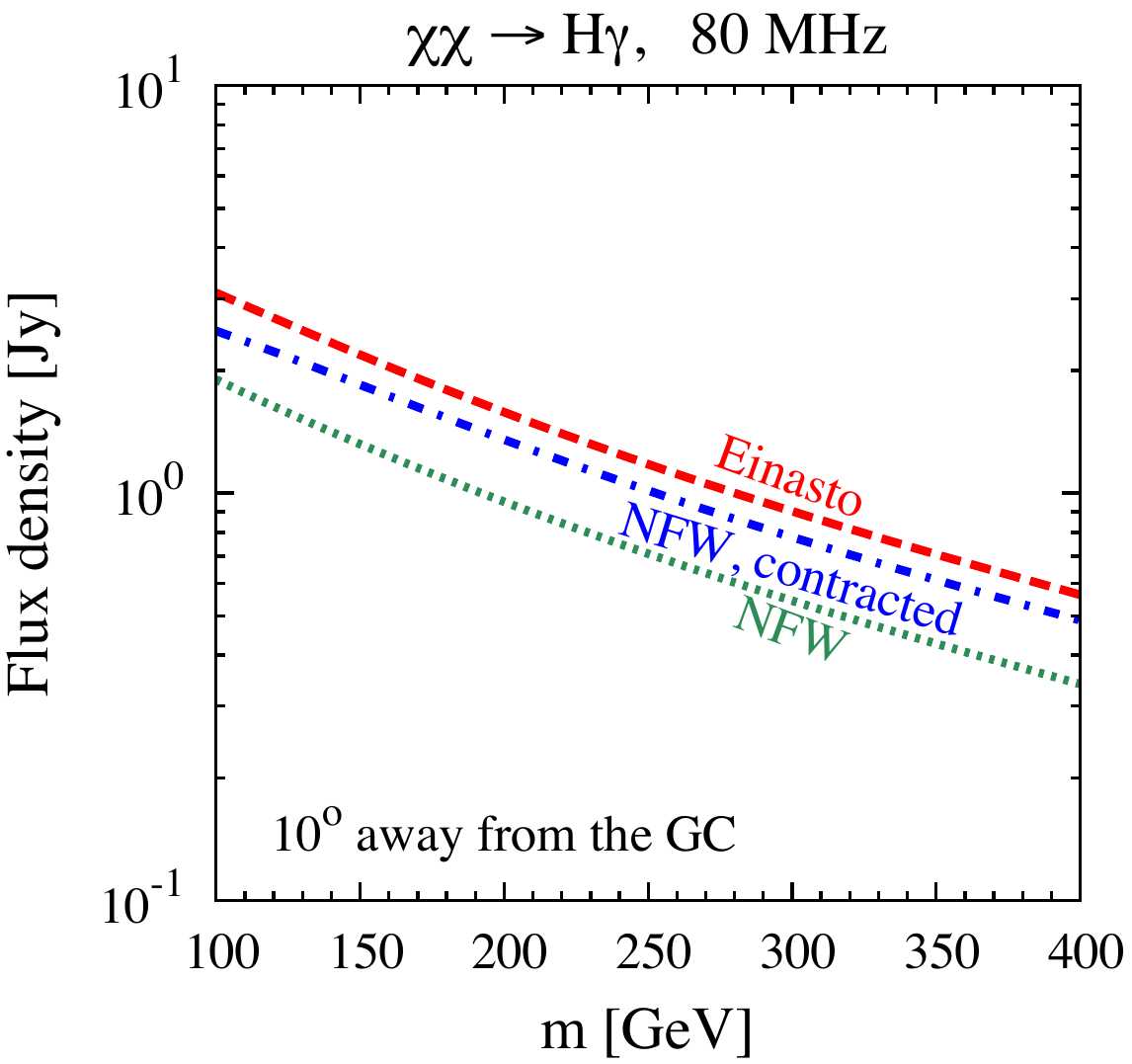}
\caption{Prediction of the synchrotron flux density at a region if radius 1$^\circ$ at $10^\circ$ away from the GC, against mass of the DM. The DM annihilation cross section and the profiles are the same as in Fig.~\ref{fig:flux density ROI2}. ({\bf Top}) Results for the 200\,MHz radio band. ({\bf Bottom}) Results for the 80\,MHz radio band. We use the exponential magnetic fields in Eq.\,(\ref{eq:large scale galactic magnetic field}).}
\label{fig:flux density future away}
\end{figure*}

%\begin{figure*}[!thbp]
%\centering
%\includegraphics[angle=0.0,width=0.45\textwidth]{I_vs_mass_200MHz_Zgamma_10power-26_FornengoB_6muG_10degree_away_from_GC_and_in_an_angular_cone_of_1degree.pdf}
%\includegraphics[angle=0.0,width=0.45\textwidth]{I_vs_mass_200MHz_Hgamma_10power-26_FornengoB_6muG_10degree_away_from_GC_and_in_an_angular_cone_of_1degree.pdf}
%\includegraphics[angle=0.0,width=0.45\textwidth]{I_vs_mass_80MHz_Zgamma_10power-26_FornengoB_6muG_10degree_away_from_GC_and_in_an_angular_cone_of_1degree.pdf}
%\includegraphics[angle=0.0,width=0.45\textwidth]{I_vs_mass_80MHz_Hgamma_10power-26_FornengoB_6muG_10degree_away_from_GC_and_in_an_angular_cone_of_1degree.pdf}
%\caption{Prediction of the synchrotron flux density against mass of the DM.  In all the plots, we set as a benchmark DM annihilation cross section $\langle \sigma v \rangle=10^{-26}$ cm$^3$ s$^{-1}$, and consider three different DM profiles: Einasto profile in~Eq.\,(\ref{eq:Einasto}), NFW profile in~Eq.\,(\ref{eq:NFW}) and the contracted NFW profile in~Eq.\,(\ref{eq:contracted NFW}). ({\bf Left}) $\chi \chi \rightarrow Z \gamma$. ({\bf Right}) $\chi \chi \rightarrow H \gamma$. ({\bf Top}) Results for 200\,MHz using the exponential magnetic fields in Eq.\,(\ref{eq:large scale galactic magnetic field}). ({\bf Bottom}) Results for 80\,MHz using the exponential magnetic field in Eq.\,(\ref{eq:large scale galactic magnetic field}). The region of interest is a circular region of radius 1$^\circ$ at an angle 10$^\circ$ away from the GC.}
%\label{fig:flux density future away 10}
%\end{figure*}

\begin{figure}[!ts]
\includegraphics[width=0.43\textwidth]{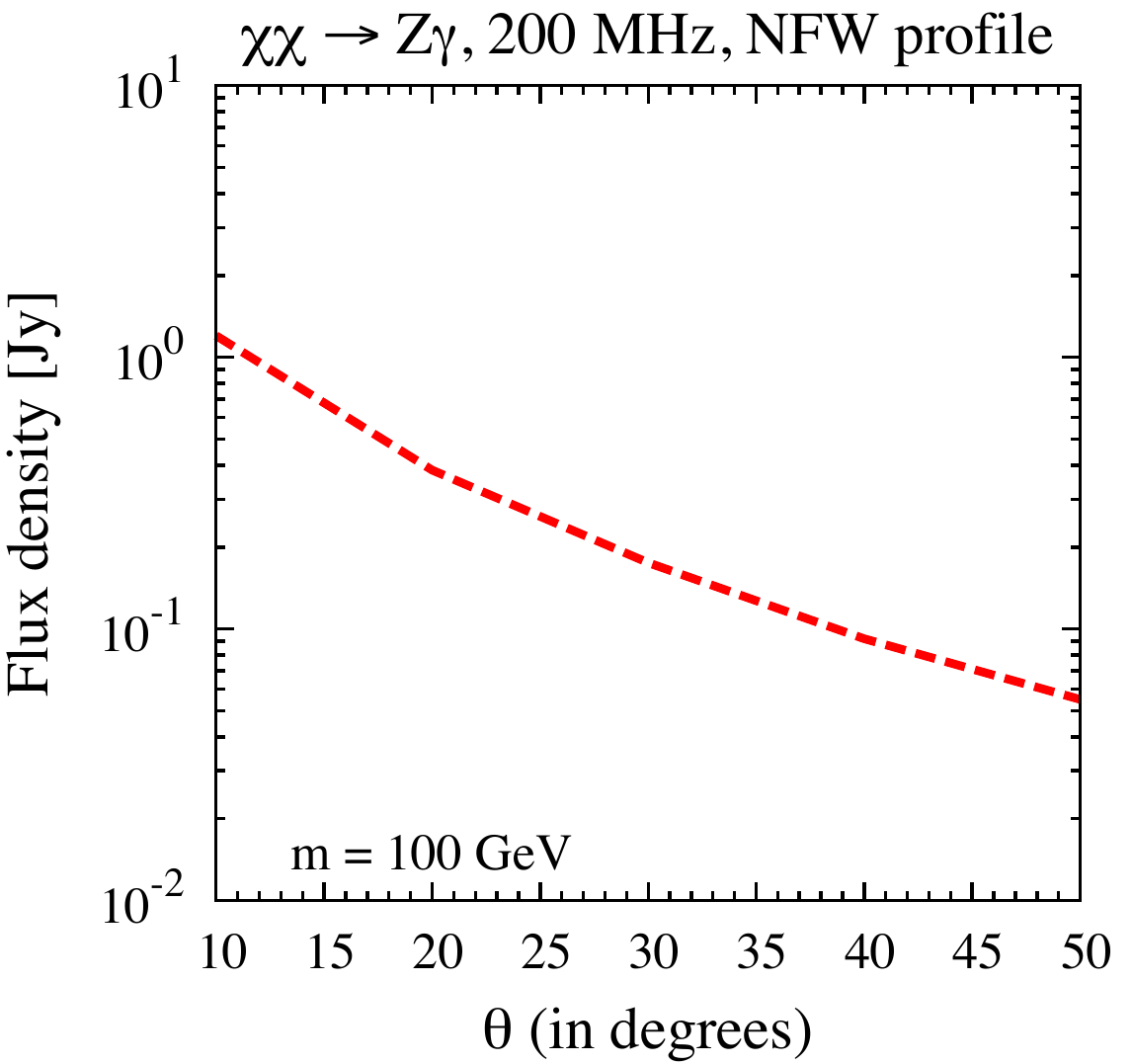}
\caption{Prediction of the synchrotron flux density against angle away from the GC for a DM mass of 100\,GeV at 200\,MHz.  The region of interest is a circular region of radius 1$^\circ$ at the specified angle away from the GC. We take the NFW DM profile, the same annihilation cross-section, and magnetic field as in Fig.~\ref{fig:flux density future away}. The variation of the synchrotron flux density with angle is very similar for all the DM profiles and annihilation channels considered in this work.}
\label{fig:I vs theta}
\end{figure}

%%%%%%%%%%%%%%%%%%%%%%%%%%%%%%%%%%%%%%%%%%
\section{Results}	\label{sec:results}
%%%%%%%%%%%%%%%%%%%%%%%%%%%%%%%%%%%%%%%%%%
In this section, we first present our results for the synchrotron fluxes from DM annihilation products, and discuss the expected systematic uncertainty due to incomplete knowledge of the DM and magnetic field profiles. 
In all the plots, we have taken $\langle \sigma v \rangle =10^{-26}\,{\rm cm}^3\,{\rm s}^{-1}$ unless otherwise mentioned. We only assume the DM self annihilation channels $\chi \chi \rightarrow Z\gamma$ or $\chi \chi \rightarrow H\gamma$ to present a completely particle physics model independent result. We then compare the expected flux with available/projected radio data from the GC to arrive at constraints/sensitivities on the DM annihilation cross section. We perform this exercise for three frequency bands (330 MHz, 408\,MHz, and 80\,MHz) in two different regions of interest around the GC. We also calculate the synchrotron flux due to DM annihilation using the above mentioned parameters in a region of radius 1$^\circ$ at an angle 10$^\circ$ away from the GC. Although we do not use the region offset to the GC to derive any constraints on DM properties, we remind the reader that strong constraints can also be obtained from radio observation at a region away from the GC.

\subsection{Synchrotron flux density at the GC}

\subsubsection{Flux density in ROI-2$^\circ$}
We calculated the synchrotron flux according to the prescription and inputs presented in Sec.\,{\ref{sec:Inputs}} and plot, in the top and bottom panels of Fig.\,\ref{fig:flux density ROI2}, the flux density due to synchrotron radiation from DM self annihilation products against mass of the DM for a region of radius 2$^\circ$ around the GC in the 330\,MHz and 80\,MHz radio band respectively. The magnetic field used is a spherically symmetric exponential magnetic field, taken from Eq.\,(\ref{eq:large scale galactic magnetic field}) with the magnetic field at the solar radius normalized to be 6\,$\mu$G. The uncertainty in the measurement of the synchrotron flux density at 330\,MHz band in this region around the GC at this frequency band, 900 Jy~\cite{2010Natur.463...65C}, is also shown in the plot and is used to obtain our constraints on the DM particle properties. 

For both the frequency bands, the synchrotron flux from DM annihilation products is maximum for the contracted NFW profile. This is expected because the signal is proportional to $\rho^2$ and a larger $\rho$ increases the signal at the GC. When the region of interest is fairly large, e.g., ROI-2$^\circ$, the Einasto profile is predicted to lead to more annihilation than the standard NFW profile. At such a large distance from the GC, the synchrotron energy density only varies by a factor of a few for different DM profiles, demonstrating the relative robustness of these results.

\subsubsection{Flux density in ROI-4$''$}

We calculated the flux densities and plot it against the mass of the DM in the top and bottom rows in Fig.\,\ref{fig:flux density ROI4} for the 408\,MHz and 200\,MHz radio band respectively. The changes are that the region of interest is now a circular region of radius 4$''$ around the GC and the frequency of radio observations is taken to be 408\,MHz. The magnetic field used is the equipartition magnetic field, taken from Eq.\,(\ref{eq:equipartition magnetic field}). We get very similar results (differences of less than 1\,mJy) if we use the cored magnetic field, as given in Eq.\,(\ref{eq:cored magnetic field}). A different value of the magnetic field at the solar radius (within the range 3\,$\mu$G to 10\,$\mu$G) does not change the value of the synchrotron flux density by more than a factor of two. The upper limit on the synchrotron flux density in this region around the GC at this frequency, 50 mJy~\cite{1976MNRAS.177..319D}, is also shown in the top plot and is used to obtain our constraints on the DM particle properties.

As expected, the synchrotron flux from DM annihilation products is maximum for the contracted NFW profile. However, in contrast to the above we find that for smaller regions of interest, e.g., ROI-4$''$ the cuspiness of NFW profile at lower radii leads to larger fluxes than from the Einasto profile. Note however, that for such small regions of interest around the GC, the flux varies by orders of magnitude for the different DM profiles.
 
\subsubsection{Flux density in ROI-away}
\label{sec:ROI-away}

We calculated the synchrotron flux according to the prescription and inputs given in Sec.~\ref{sec:Inputs} in ROI-away and plot some representative results in Fig.~\ref{fig:flux density future away}. The top and bottom panels show the synchrotron flux in a circular region of radius 1$^\circ$ at 10$^\circ$ away from the GC for the 200\,MHz and 80\,MHz radio band respectively. The magnetic field used is the exponential magnetic field, taken from Eq.\,(\ref{eq:large scale galactic magnetic field}) with the magnetic field at the solar radius normalized to be 6\,$\mu$G. We also take into account the variation of the radiation density with distance from the GC while calculating the synchrotron fluxes following the parametrization in Eq.\,(\ref{eq:radiation density}). For a given angle $\theta$ away from the GC, we calculate the value of $z$ and then use the radiation energy density $U_{\rm stellar}(0,z)$ in our calculations. Although this is a conservative approximation, we expect that a full calculation will given similar results.

The disadvantage of the synchrotron flux decreasing as one makes a measurement away from the GC is mitigated by the fact that the flux depends less strongly on the assumed DM profile. Due to the excellent sensitivity of present generation radio telescopes like LWA and LOFAR and even better sensitivity of near future radio telescopes like SKA very robust limits on DM properties can be obtained  from radio measurements away from the GC. In particular, as mentioned earlier, if the measurement is done in a radio cold spot then modeling the astrophysical backgrounds will also be easier to find the putative radio signal of DM annihilation.

\subsubsection{Variation of the synchrotron flux with angle away from the GC}

We plot our calculated synchrotron flux density against angle away from the GC  for the 200\,MHz band and for the $\chi \chi \rightarrow Z\gamma$ channel in Fig.~\ref{fig:I vs theta}. We assume the exponential magnetic field as given in Eq.\,(\ref{eq:large scale galactic magnetic field}). We calculate our synchrotron fluxes in a region of radius 1$^\circ$ at the specified angle away from the GC. We take into account the variation in the radiation field energy density following the prescription in Sec.~\ref{sec:ROI-away}. As can be seen from the plots, the synchrotron flux falls off by an order of magnitude as the angle away from the GC increases from 10$^\circ$ to 50$^\circ$. A very similar variation of the synchrotron flux away from the GC is obtained for the $\chi \chi \rightarrow H\gamma$ channel and in the 80\,MHz radio band.

\subsection{Sensitivity to magnetic fields}

Now, we explore the sensitivity of the predicted synchrotron fluxes to the normalization and shape of the Galactic magnetic field profile. We remind the reader that we have used 6\,$\mu$G as our benchmark value of the magnetic field at the solar radius for all our calculations presented in the other sections. The DM is assumed to have a standard NFW profile (Eq.\,(\ref{eq:NFW})) for the plot in this section.

\begin{figure}[!t]
\centering
\vspace{-0.1cm}
\includegraphics[angle=0.0, width=0.427\textwidth]
{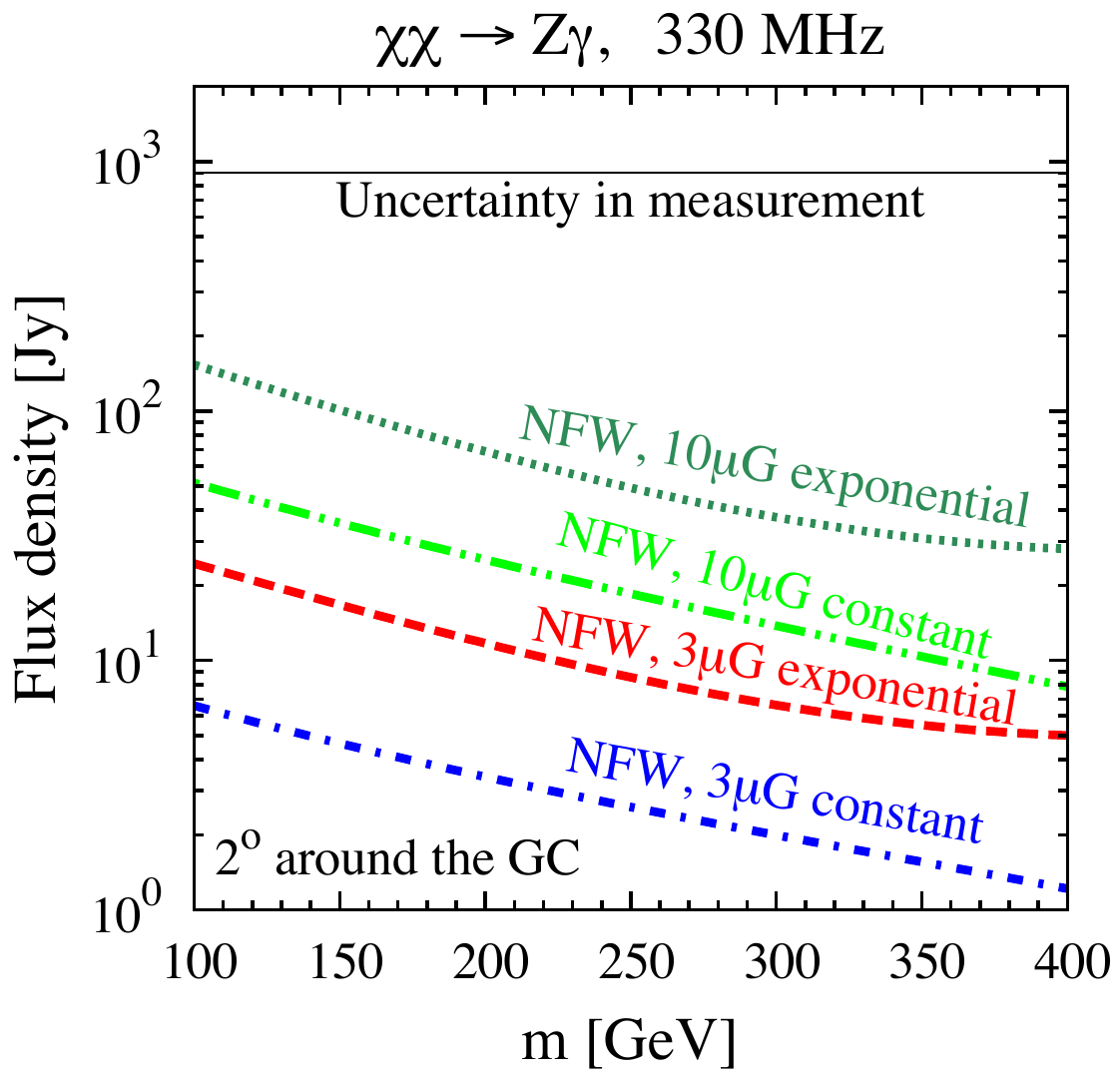}
\caption{Prediction of the synchrotron flux density vs. mass of the DM at 330\,MHz, in a region 2$^\circ$ around the GC. The magnetic field used is exponential, as in Eq.\,(\ref{eq:large scale galactic magnetic field}), and a constant magnetic field. We use two different values of the magnetic fields at the solar radius: 3\,$\mu$G and 10\,$\mu$G. We use the NFW profile and the same annihilation cross section as in Fig.~\ref{fig:flux density ROI2}. The variation is similar for all the DM profiles and annihilation channels considered in the text.}
\label{fig:flux density for different values of the solar magnetic field}
\end{figure}

To understand the impact of the normalization of the Galactic magnetic field on the synchrotron flux density due to dark matter annihilation, in Fig.\,\ref{fig:flux density for different values of the solar magnetic field} we plot the synchrotron flux due to two different values of the Galactic magnetic field at the solar radius: 3$\,\mu$G and 10$\,\mu$G. It is seen that varying the normalization of the Galactic magnetic field can change the synchrotron flux density by a factor of a few for both the exponential magnetic field profile and the constant magnetic field profile.

We show the impact of two different magnetic field profiles: the exponential profile, Eq.\,(\ref{eq:large scale galactic magnetic field}), and the constant magnetic field profile for the 330 MHz band. For a given normalization of the magnetic field profile at the solar radius, we see that the flux due to the exponential magnetic field profile is always larger than the flux due to the constant magnetic field profile for the DM annihilation channel, $\chi \chi \rightarrow Z\gamma$. The result is similar for the annihilation channel $\chi \chi \rightarrow H\gamma$.

The overall uncertainty in the normalization and the shape of the Galactic magnetic field can lead to a difference of at most an order of magnitude in the predicted synchrotron flux in the 2$^\circ$ around the GC at 330\,MHz. We have checked that the variation in the synchrotron flux density with the normalization and shape of the Galactic magnetic field is similar for the other DM profiles considered in this work. 

For the region of angular radius 4$''$ around the Galactic Center, the difference in the synchrotron flux density is less than a factor of two for both the equipartition, Eq.\,(\ref{eq:equipartition magnetic field}) and the cored magnetic field profile, Eq.\,(\ref{eq:cored magnetic field}), for a given DM profile. Again, the uncertainty due to incomplete knowledge of magnetic fields can lead to at most an order of magnitude changes in the predicted synchrotron fluxes.

\subsection{Constraints on $\sigv$-$m$}

\begin{figure*}[!htbp]
\centering
\includegraphics[angle=0.0, width=0.44\textwidth]{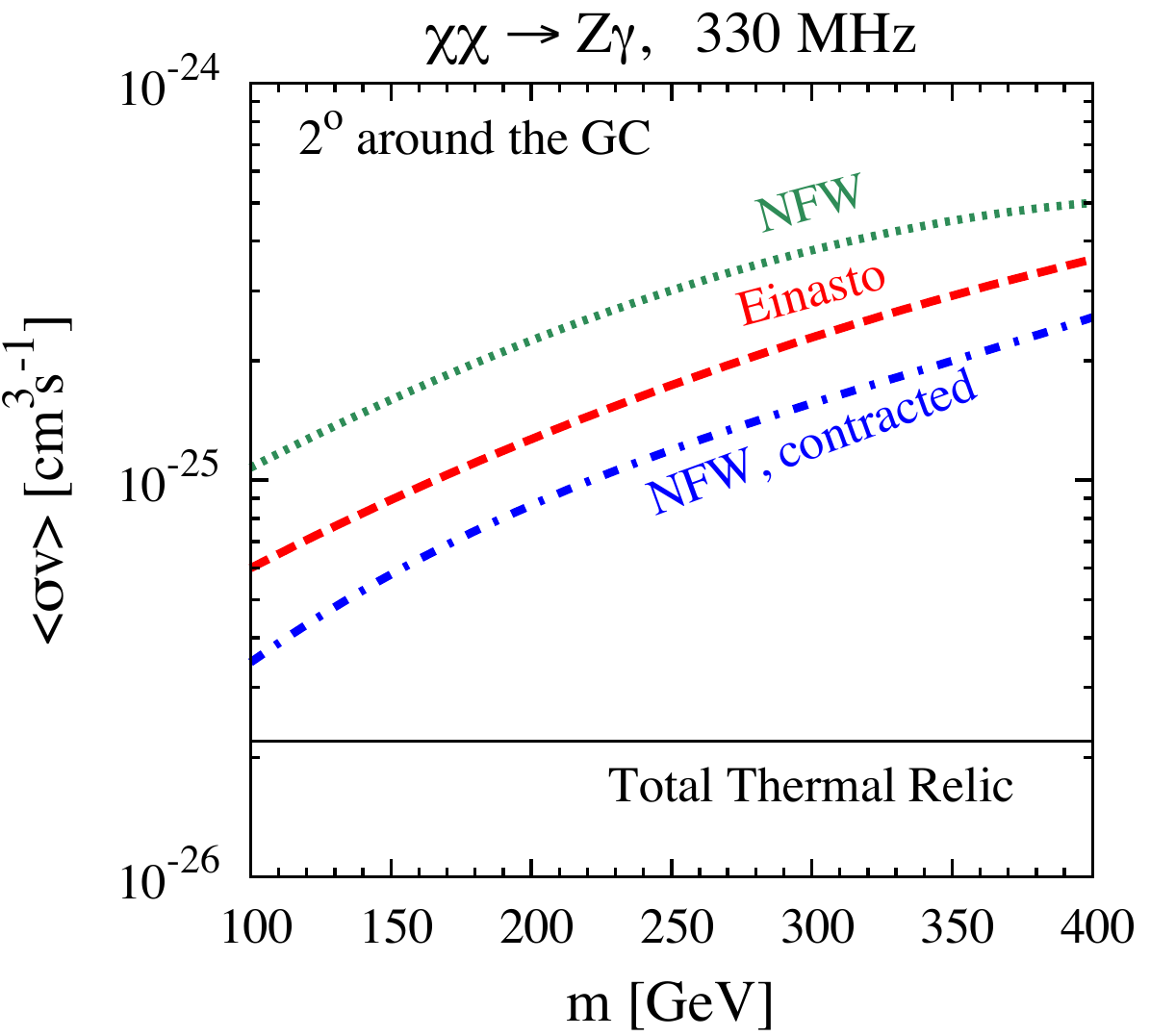}
\includegraphics[angle=0.0, width=0.44\textwidth]{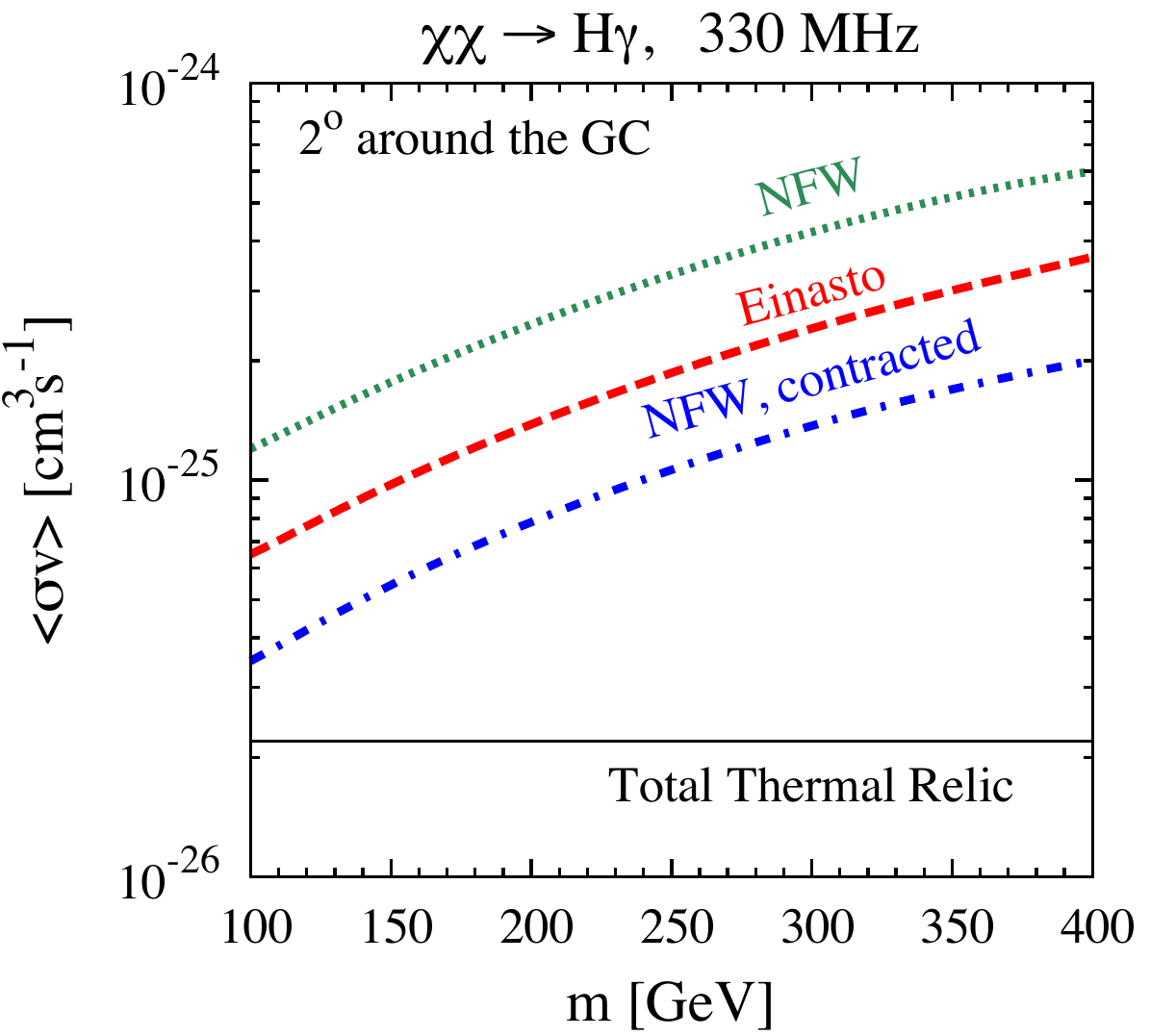}

\includegraphics[angle=0.0,width=0.44\textwidth]{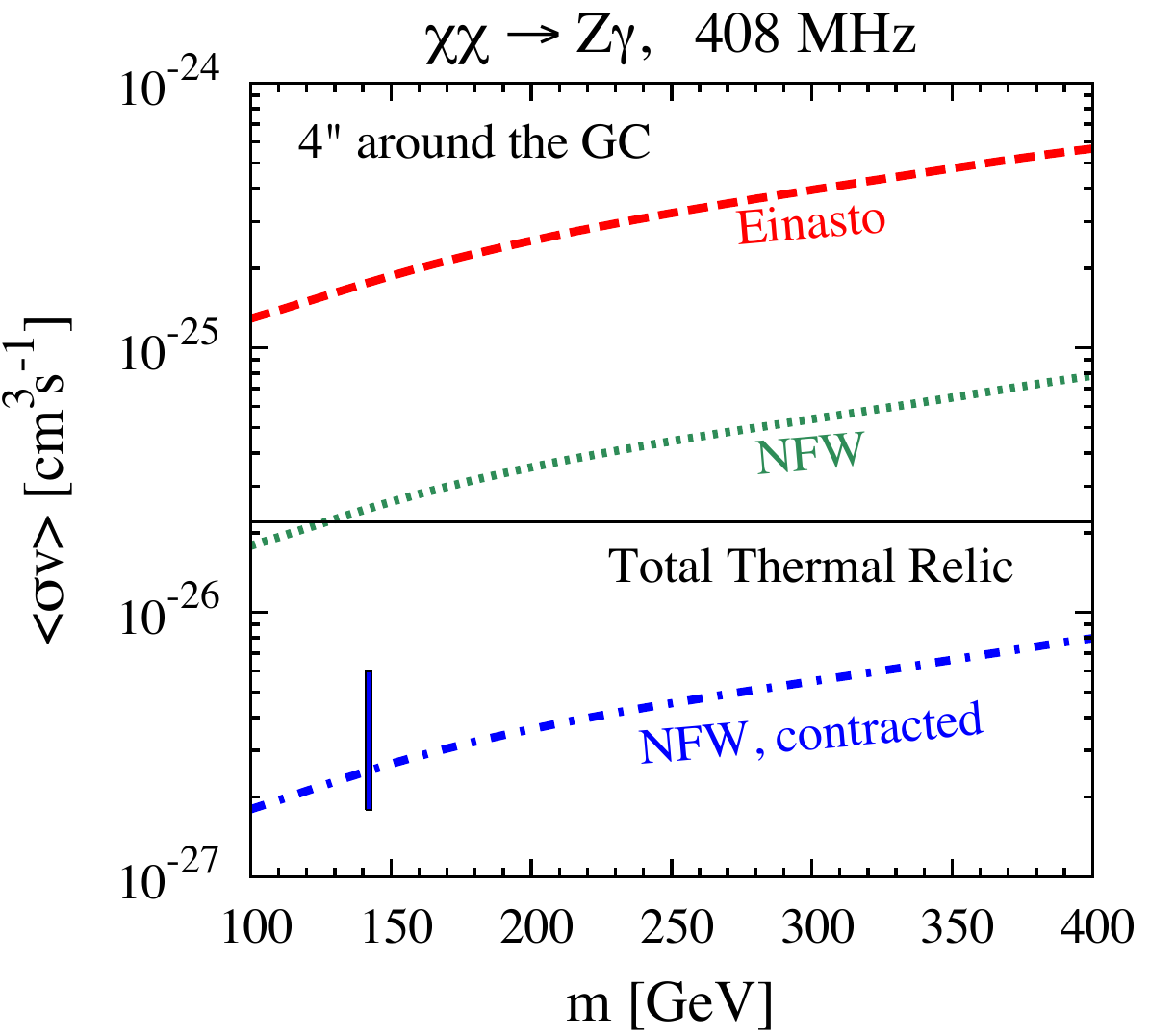}
\includegraphics[angle=0.0,width=0.44\textwidth]{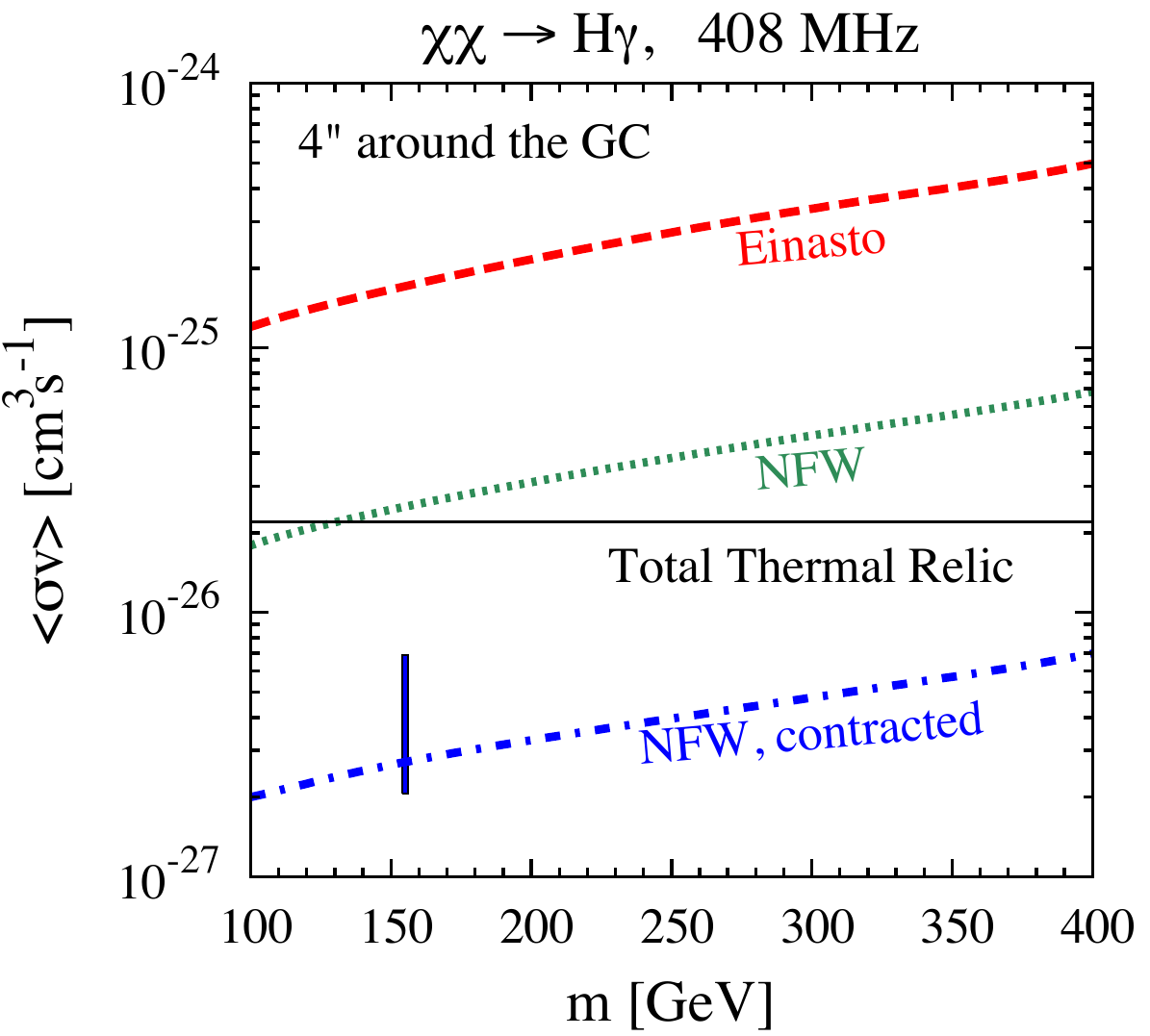}

\includegraphics[angle=0.0,width=0.44\textwidth]{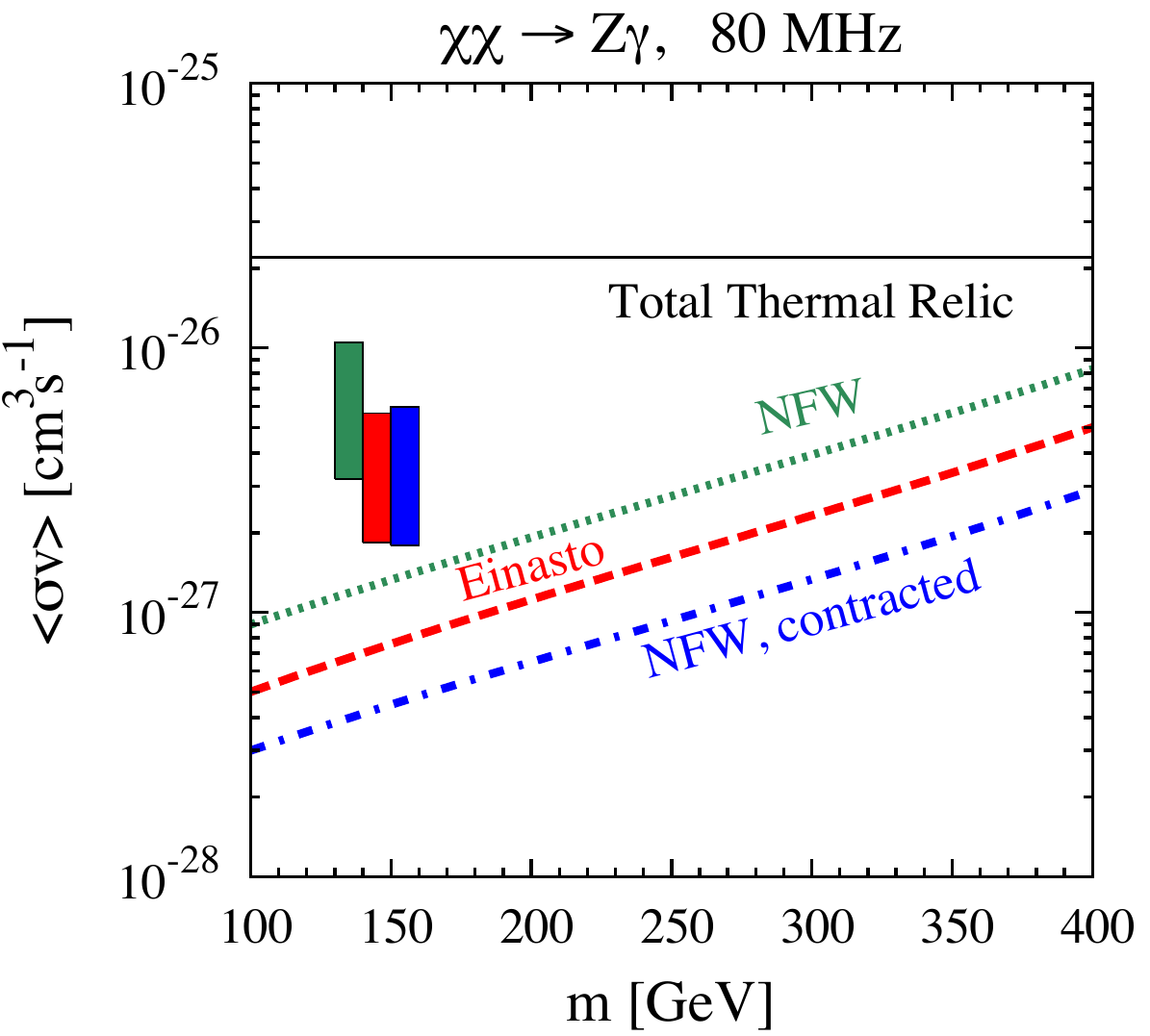}
\includegraphics[angle=0.0,width=0.44\textwidth]{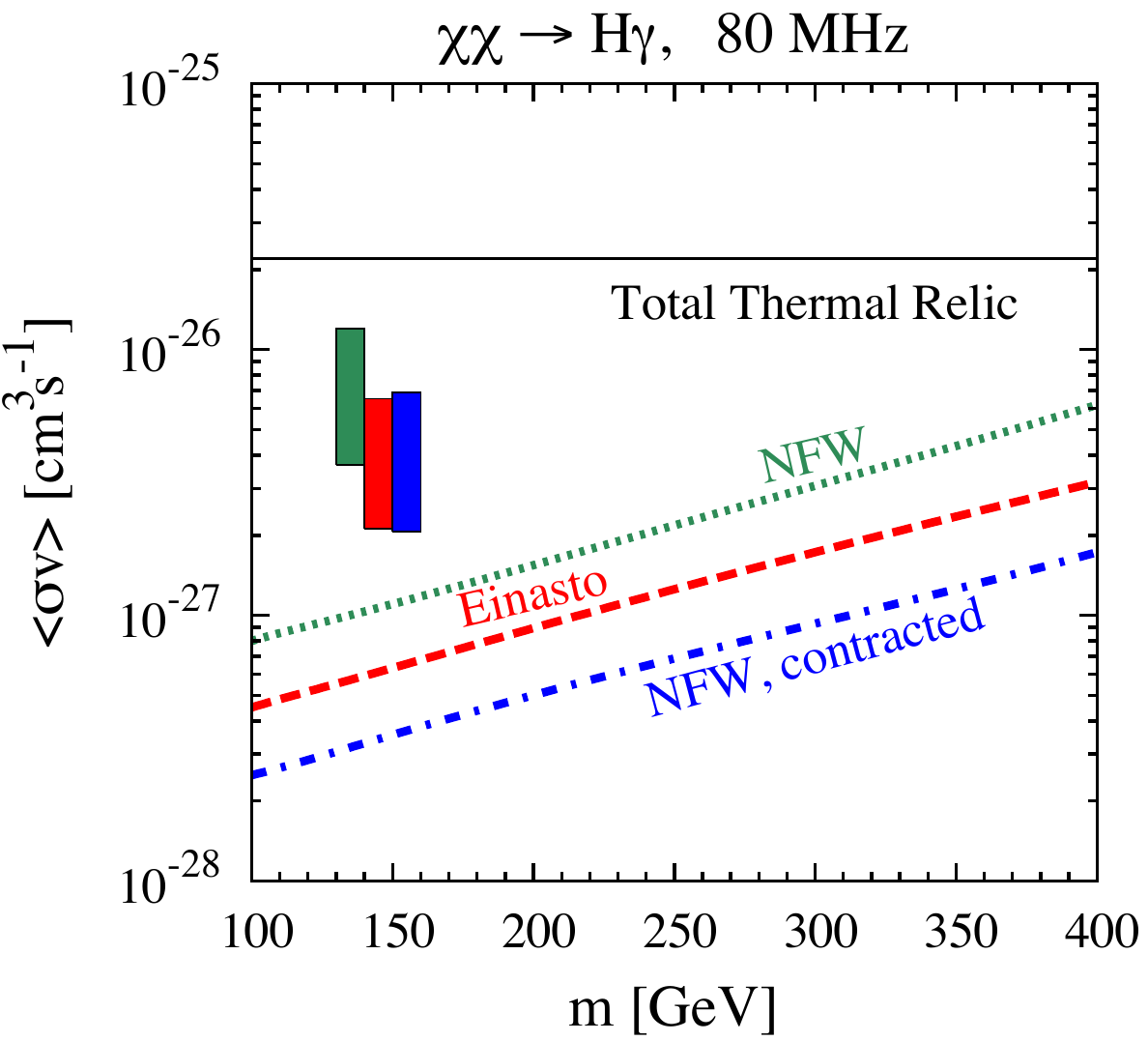}
\caption{Constraints obtained in the $\sigv$ vs. {\it m} plane for different annihilation channels, different frequency bands, and different regions of observations. The DM profiles are the same as in Fig.~\ref{fig:flux density ROI2}. ({\bf Left}) $\chi \chi \rightarrow Z \gamma$. ({\bf Right}) $\chi \chi \rightarrow H \gamma$. ({\bf Top}) Results for 330\,MHz using the exponential magnetic field in Eq.\,(\ref{eq:large scale galactic magnetic field}). ({\bf Middle}) Results for 408\,MHz using the equipartition magnetic field in Eq.\,(\ref{eq:equipartition magnetic field}). The DM mass and $\langle \sigma v \rangle$ preferred for the 130 GeV gamma-ray line is shown by the thin box (see text for more details). ({\bf Bottom}) The sensitivities at 80\,MHz using the exponential magnetic field in Eq.\,(\ref{eq:large scale galactic magnetic field}). The shaded region shows approximately the $\langle \sigma v \rangle$ favored by the recent 130\,GeV gamma-ray line signal for the NFW profile (green), Einasto profile (red) and contracted NFW profile (blue)~\cite{Weniger:2012tx}. The favored region in masses is same as the plot above, and are staggered in this bottom plot for better legibility. We also show the updated total thermal relic cross section from reference~\cite{Steigman:2012nb}.}
\label{fig:constraint}
\end{figure*}

\subsubsection{Constraint from the measurement at 330\,MHz}

In the region with radius 2$^\circ$ around the GC, data in the 330\,MHz radio band is presented in Ref.\,\cite{2010Natur.463...65C}. We compare our prediction of the synchrotron flux from products of DM annihilation, and demand that the radio is not over-saturated by the DM-induced fluxes. This gives us a constraint on the DM annihilation cross section $\sigv\sim10^{-25}\,{\rm cm^3\,s^{-1}}$. This is overly conservative, as there are known astrophysical sources that produce most of the observed synchrotron radiation. The astrophysical model presented in Ref.\,\cite{2010Natur.463...65C} suggests that with present level of uncertainty, at most 5\% of the flux ($\lesssim 900$\,Jy) could come from unknown sources. This gives a much stronger constraint $\sigv \sim10^{-26}\,{\rm cm^3\,s^{-1}}$. This constraint on the $\sigv$-$m$ plane that can be derived from the radio flux measurement at 330\,MHz for a circular region of radius 2$^\circ$ is plotted in the top panel of Fig.\,\ref{fig:constraint}. We only show the constraints that can be obtained in this radio band by using the exponential magnetic field given in Eq.\,(\ref{eq:large scale galactic magnetic field}), with a normalization of 6\,$\mu$G. 

For both the DM self annihilation channels $\chi \chi \rightarrow Z\gamma$ and $\chi \chi \rightarrow H\gamma$, we see that the contracted NFW profile gives the most constraining limit. Since the gamma-ray line prefers a cross section $\sigv\sim10^{-27}\,{\rm cm^3\,s^{-1}}$ for all the three profiles~\cite{Weniger:2012tx}, it can be concluded that the existing data at this frequency is not able to constrain the line signal independent of a DM particle physics model. However, since the present constraints are only an order of magnitude away from the DM self annihilation cross section preferred by the 130\,GeV signal, a future radio measurement near the GC can be used to either constrain or confirm its presence at the GC.

\subsubsection{Constraint from the measurement at 408\,MHz}

The upper-limit on the synchrotron flux at 408\,MHz found by Ref.\,\cite{1976MNRAS.177..319D} allows us to impose much stronger constraints than above. The procedure that we follow is similar to above - we compare the predicted fluxes with the existing upper limit, and demand that the DM annihilation not produce a flux larger than what is already constrained. This constraint in the $\sigv$-$m$ plane that can be derived from the radio flux measurement at 408\,MHz for a circular region of radius 4$''$ is plotted in the middle row of Fig.\,\ref{fig:constraint}. 

We also show the dark matter mass and self-annihilation cross section preferred for the 130 GeV gamma-ray line by the thin shaded box. For the annihilation to $Z\gamma$/ $H\gamma$, the gamma-ray energy is given by $E_{\gamma}=m_{\chi}\left(1-m_{Z/h}^2/4\,m_{\chi}^2 \right)$. Hence for a 130\,GeV gamma-ray line, the DM mass preferred is $\sim$142\,GeV for annihilation to $Z\gamma$ and a DM mass of $\sim$155\,GeV is preferred for annihilation to $H\gamma$. Given the self annihilation cross section $\langle \sigma v \rangle _{\gamma \gamma}$ presented in Ref.\,\cite{Weniger:2012tx}, we convert them to $\langle \sigma v \rangle _{Z/H \gamma}$ by following the prescription given in Ref.\,\cite{Vertongen:2011mu}. For DM self-annihilation to $Z\gamma$ or $H\gamma$, the relation between DM mass and the gamma-ray line is given by $m_{\chi}=(1/2)(1+\sqrt{1+m_{Z/h}^2/E_{\gamma}^2})E_{\gamma}$, and it follows from kinematic considerations that if the limits of $\langle \sigma v \rangle _{\gamma \gamma}$ are given, the corresponding limits for $\langle \sigma v \rangle _{Z/H \gamma}$ is given by $\langle \sigma v \rangle _{Z/H \gamma}=(1/2)\left(1+\sqrt{1+m_{Z/h}^2/E_{\gamma}^2}\right)^2 \langle \sigma v \rangle _{\gamma \gamma}$. We take the upper and lower limits on $\langle \sigma v \rangle _{\gamma \gamma}$ for the 130\,GeV gamma-ray line from the Region 4 of the SOURCE class events as presented in Ref.\,\cite{Weniger:2012tx}. Using the other regions and the ULTRACLEAN class events gives similar limits and it will not change our conclusions. For the 408\,MHz radio band, we only show the $\langle \sigma v \rangle _{Z/H \gamma}$ that is preferred by the 130\,GeV gamma-ray line for the NFW contracted profile.

For both the DM self annihilation channels $\chi \chi \rightarrow Z\gamma$ and $\chi \chi \rightarrow H\gamma$, we see that the contracted NFW profile gives the most constraining limit ($\sigv\lesssim 10^{-27}\,{\rm cm^3\,s^{-1}}$), and in fact the sensitivity to the cross section is less than the total thermal relic cross section for both the self annihilation channels. The least constraining limit is obtained from the Einasto DM profile, as expected ($\sigv\lesssim 10^{-25}\,{\rm cm^3\,s^{-1}}$). If we assume that the modeling of the magnetic field near the GC black hole is correct, then this shows that the interpretation of the line signal near the GC for a contracted NFW profile is in mild tension with the radio data, provided the source of the gamma-ray line in the GC is due to the $\chi \chi \rightarrow Z\gamma$ and $\chi \chi \rightarrow H\gamma$ self annihilation channel.

\subsubsection{Sensitivity from a future measurement at 80\,MHz}

The situation is expected to improve dramatically with future observation of the GC by LWA, LOFAR, and SKA. Although we present our future constraint from a radio flux measurement at 80\,MHz near the GC, it is worth mentioning that strong constraints can also be obtained from measurement of the radio flux away from the GC. As mentioned earlier, ideally we expect the best measurement to come from a radio cold spot. The standard astrophysical background has to be modeled very carefully to reach the sensitivity as presented in this paper.

 To forecast the sensitivity, we very conservatively assume that LWA can reach a background subtracted flux density sensitivity 10\,Jy at 80\,MHz for a circular region of radius 2$^\circ$ around the GC~\cite{Kassim}.
The constraint in the $\sigv$-$m$ plane that can be derived from the radio flux measurement at 80\,MHz for a circular region of radius 2$^\circ$ is plotted the bottom panel of Fig.\,\ref{fig:constraint}. We also show the $\langle \sigma v \rangle _{Z/H \gamma}$ preferred by the 130\,GeV gamma-ray line by the green, red and blue shaded boxes for the NFW, Einasto and the contracted NFW DM profile respectively. We again use the Region 4 in the SOURCE class events and the prescription given in the previous section to draw these boxes. We did not draw these shaded boxes in the correct DM mass positions for clarity. 

Due to the superior flux sensitivity of LWA at these frequencies, we see that both the $\chi \chi \rightarrow Z\gamma$ and $\chi \chi \rightarrow H\gamma$ self annihilation channel can be probed well below the total thermal relic cross section for all three considered DM profiles. In particular, for all the DM profiles considered, one can probe below the $\sigv\sim 10^{-27}\,{\rm cm^3\,s^{-1}}$ cross sections required to explain the tentative 130\,GeV signal. Thus, if the 130\,GeV gamma-ray line turns out to be robust and originates from DM self annihilation, LWA has a good chance to search for the self annihilation channel giving rise to the line for the NFW, Einasto and the contracted NFW profile. Up to the uncertainty in the GC model, this remains, to our knowledge, the best probe for discerning the origin of the DM line independent of any particle physics DM model. Since LWA will reach this sensitivity over a large region of observation, the dependence of the constraint on the underlying DM profile is modest. We expect similar limits can be obtained by the LOFAR collaboration as well. SKA is expected to further strengthen this constraint.

%%%%%%%%%%%%%%%%%%%%%%%%%%%%%%%%%%%%%%%%%% 
\section{Summary and Outlook}
\label{sec:outlook}
%%%%%%%%%%%%%%%%%%%%%%%%%%%%%%%%%%%%%%%%%%
In this paper we have shown that existing radio data around the Galactic Center at 408\,MHz marginally constrains the interpretation of the 130\,GeV line in Fermi-LAT data in terms of DM self annihilation to $Z\gamma$ or $H\gamma$ with a cross section $\sim$ $\,10^{-27}\,{\rm cm^3\,s^{-1}}$ for a contracted NFW profile. For other frequencies or other DM density profiles the constraint is up to an order of magnitude weaker within the parameter ranges chosen by us. Future measurements made around the GC by LWA in the 80\,MHz band can push the sensitivity to DM annihilating to gamma-ray lines down to $\sigv\sim10^{-28}\,{\rm cm^3\,s^{-1}}$ and enable a test of the above signal. Although the background needs to be known very well to achieve our quoted limits, these possibilities are, to the best of our knowledge, some of the most competitive ways to test for the nature of the DM that could have produced the tentative 130\,GeV line signal.

We have shown that these conclusions are fairly robust with respect to the assumptions on the magnetic field in the Galaxy, and the constraints do not weaken by more than an order of magnitude. The dependence on DM density profiles is somewhat more important, especially when the region of observation is small and closely centered on the GC. While the uncertainty in the astrophysical modeling of the GC does impact our results (see for e.g., \cite{Everett:2007dw} for a different modeling of the GC), we must emphasize that these constraints are completely model-independent from the particle physics perspective, because we have simply taken the electrons and positrons from the known decays of the $Z$ or $H$ produced in the DM annihilation to $Z\gamma$ or $H\gamma$, respectively. A similar study on dark matter annihilation contribution to the galactic radio background~\cite{Kogut:2009xv} and diffuse extragalactic radio background~\cite{Fixsen:2004hp,Seiffert:2009xs} can also performed to cross-check potential dark matter signals from the Galactic Center~\cite{Hooper:2012jc,Fornengo:2011cn,Fornengo:2011xk}.

We hope that these results will encourage radio astronomers, especially those at LWA, VLA-Low, LOFAR, and SKA, to observe the GC, model the astrophysical synchrotron backgrounds, and determine if there is any excess flux. Irrespective of whether the tentative 130\,GeV gamma-ray line signal at Fermi-LAT is due to DM annihilation or not, this promises to deliver some of the strongest constraints on DM annihilation.

\section*{Acknowledgments}
We thank John Beacom, Sayan Chakraborti, Christopher Kochanek, Kohta Murase, Brian Lacki, and Todd Thompson for helpful discussions. B.D. and R.L. also thank Ilias Cholis, Alex Geringer-Sameth, Joachim Kopp, Tim Linden, Marco Taoso, Douglas Spolyar, and Christoph Weniger for interesting questions and suggestions at the IDM 2012 conference where this work was first presented. We are especially grateful to Namir Kassim and Joseph Lazio who informed us about LWA and whose advice greatly improved the paper. R.\,L. is supported by NSF Grant PHY-1101216 awarded to John Beacom.

\vspace{0cm}
\bibliographystyle{kp}
\interlinepenalty=10000
\tolerance=100
\bibliography{Bibliography/references}

\end{document}